\documentclass[a4paper,11pt]{article}
\pdfoutput=1 % if your are submitting a pdflatex (i.e. if you have
\usepackage{jheppub} % for details on the use of the package, please
\usepackage[T1]{fontenc} % if needed

\usepackage{pdfsync}
\usepackage{mathptmx}
\usepackage[utf8]{inputenc}
\usepackage[T1]{fontenc}
\usepackage{slashed}
\usepackage{times}
\usepackage{amssymb,amsfonts,amsmath,amsthm}
\usepackage{dsfont,bbm}
\usepackage{dcolumn}
\usepackage{epsf}
\usepackage{graphicx}
\usepackage[caption=false]{subfig}
\usepackage{dsfont}
\usepackage{bm}
\usepackage{slashed}
\usepackage[active]{srcltx}
\usepackage[usenames]{color}
\newcommand{\up}{{\uparrow}}

\title{\boldmath Archetypal Factorization and Gluon Poles in semi-exclusive reactions}

\author[a,b]{I.~V.~Anikin,}
\author[b]{L.~Szymanowski}

\affiliation[a]{Bogoliubov Laboratory of Theoretical Physics, JINR,\\141980 Dubna, Russia}
\affiliation[b]{National Centre for Nuclear Research (NCBJ),\\02-093 Warsaw, Poland}

\emailAdd{Igor.Anikin@ncbj.gov.pl}
\emailAdd{Lech.Szymanowski@ncbj.gov.pl}

\abstract{Within the archetypal factorization procedure, we study the gluon pole contributions
manifesting in the nucleon-lepton hard processes of hadron production.
We prove the dominant role of gluon pole contributions in the processes of such kind.
We analyse the different sources of complexity associated with the gluon pole functions.
As a practical application, we derive a new single spin asymmetry
related directly to the gluon poles which can be studied experimentally.}

\begin{document}
\maketitle
\flushbottom

%%%%%%%%%%%%%%%%%%%%%%%%%%%%%
\section{Introduction}
\label{Sec:Int}
%%%%%%%%%%%%%%%%%%%%%%%%%%%%%

The hadron production in nucleon-lepton processes attracts still much attentions
not only of the experimental collaborations (for example, at JLab)
which investigate the composite (spin) structure of hadrons.
From the theoretical viewpoint, this kind of processes is interested because
it opens the window for the studies of the transverse momentum dependent functions,
in particular, appearing in the frame of semi-inclusive kinematics, see Fig.~\ref{Fig-S-1}.
In this kinematics, the hadron tensor involves both the distribution and
fragmentation functions which are linked by the
loop integrations over the transverse components of momenta (this is a general feature of TMD-factorizations).
Based on the classical formulation of the factorization theorem
\cite{Efremov:1976ih,  Efremov:1978rn, Efremov:1978xm, Efremov:1978cu},
the different non-perturbative functions (such as distribution functions and fragmentation functions)
bound by the dimensionful transverse momentum integration
cannot be considered as totally independent ones.
Nevertheless, there are different approaches
where the mentioned non-factorized effects can be diminished by the
corresponding summation taking into account the transverse momenta of partons
(see, for example, \cite{Barone:2001sp} and the references therein).
Moreover, in the framework of semi-inclusive kinematics, the photon and produced hadron
transverse momenta are related each other with the necessary condition
$\vec{\bf q}_\perp = \vec{\bf P}^{h}_\perp/z \sim \vec{\bf k}_\perp$,
where $\vec{\bf k}_\perp$ denotes the primordial transverse momentum of quark.

On the other hand, the attempts to use the approaches inspired by TMD-factorization
for description of the existing experimental data from JLab are encountered the difficulties
at rather a large value of $\vec{\bf q}_\perp$ (sometimes, it is named as the ``$q_T$ crisis'').

However, the hadron production in nucleon-lepton collision can be also considered using the
semi-exclusive kinematics where the hadron tensor contains two distribution amplitudes instead of
one fragmentation function, see Fig.~\ref{Fig-S-1}.
In the present paper, alternatively to the semi-inclusive description,  we propose an approach based on the well-defined and
archetypal factorization procedure
(see for example \cite{Anikin:2020ipg, Anikin:2017azc, Anikin:2017pch, Anikin:2017vvn, Anikin:2018fom})
applied to the semi-exclusive mode of the hadron production processes.
Within this approach, the hadron tensor is given by the
corresponding convolution of the distribution function and the
distribution amplitudes if a value of intermediate quark fraction is large, $x\to 1$.
Another feature of the used approach is that the photon and produced hadron
transverse momenta are not tied in contrast to the approaches based on TMD-factorization.
In this context, we calculate a new single spin asymmetry which is associated with the gluon pole contribution
for a wide region of produced hadron transverse momentum.
This asymmetry can be also a object of experimental measurements
at a large value of $\vec{\bf P}^{h}_\perp$ using the existing data of JLab.

In the paper, we also give a proof that, in the limits of  $x\to 1$ and $m_q\to 0$,
the leading order hadron tensor is actually suppressed in comparison with the
gluon pole contribution to the hadron tensor where the gluon radiation has been taken into account.
Based on this our principle statement, we concentrate on the manifestation
of gluon poles in  the lepton-hadron
collision of hadron production.
%This process resembles the well-know Drell-Yan process (DY-like process)
%at large $x$ where the produced hadron is described by the corresponding distribution amplitude.

As shown in \cite{Anikin:2017azc, Anikin:2017pch, Anikin:2017vvn, Anikin:2018fom, Anikin:2021osx},
the use of contour gauge conception gives the strong mathematical evidences for the nature of functional complexity in the
treatment of gluon poles. From the viewpoint of practical uses, the complexity of the corresponding distribution
functions related to the gluon poles is very important because it leads to the different kinds of asymmetries.
In addition to the contour gauge conception, we investigate the different sources
of functional complexity which are associated with the gluon pole effects.

Within the factorization procedure we adhere in our paper,
we introduce a new kind of the transverse momentum dependent
distribution functions which leads to the different new observables.
With the help of these new functions and
the gluon pole contributions it becomes to be possible to
study the hadron spin structure in a different light using
the experimental data extracted from the measurements of processes with a large value of
transverse momentum of produced hadron.

The paper is organized as follows. The kinematics is introduced and
the Sudakov decompositions are presented in Sec.~\ref{Sec:Kin}.
In Sec.~\ref{Sec:HT-DY}, we obtain the leading order hadron tensor
and the hadron tensor which involves the gluon pole contribution.
Also, we describe the main stages of our factorization procedure
focusing on its important features.
In Sec.~\ref{Sec:Comp-LO-NLO}, we formulate the principal statement on
the dominant role of gluon pole contributions to the
corresponding hadron tensor in comparison with the
hadron tensor without the gluon pole contribution.
Sec.~\ref{Sec:GI-HT} is devoted to the derivation of
the gauge-invariant hadron tensor with gluon pole contributions
and to the introduction of the new transverse momentum dependent functions.
We also analyse the different sources of complexity for the
function involving the gluon pole.
The new single spin asymmetry is presented and discussed in Sec.~\ref{Sec:SSA} .
Sec.~\ref{Sec:Concl} is reserved for a summary and outlook.
Some technical issues are presented in appendices.
Appendix~\ref{App:Frame} contains the discussion on the different choices of the frame systems.
The alternative representation of functions with the
on-shell fields in correlators is presented in Appendix~\ref{AppA}.
The off-shell extension of the fields in the corresponding correlators is
discussed in Appendix~\ref{AppB}.  
In Appendix~\ref{NF-Par}, the extension of standard Lorentz parametrization is presented 
and Appendix~\ref{AppC} is devoted to the method of covariant integrations.

%%%%%%%%%%%%%%%%%%%%%%%%%%%%
\section{Kinematics}
\label{Sec:Kin}
%%%%%%%%%%%%%%%%%%%%%%%%%%%%

To begin with, we specify our kinematics.
We study the hadron (pion, $\rho$-meson) production in the hard process which
can be considered, from the kinematical point of view, as the intermediate one
between the semi-inclusive and semi-exclusive reactions.
Namely, we deal with
\begin{eqnarray}
\label{proc-1}
\ell(l_1) + N^{(\uparrow\downarrow)}(P_2)
\to \ell(l_2) + h(P_1^h) + \bar q(K) + X(P_X)
\end{eqnarray}
or
\begin{eqnarray}
\label{proc-2}
\gamma^*(q) + N^{(\uparrow\downarrow)}(P_2)
\to h(P_1^h) + \bar q(K) + X(P_X).
\end{eqnarray}
In Eqns.(\ref{proc-1}) and (\ref{proc-2}), $h$ denotes $\pi$ or $\rho$ triplet of mesons (in general, any hadron)
and the virtual photon produced by the lepton ($l_1=l_2+q$) has a large mass squared
($q^2=-Q^2$) while the photon transverse momenta are small.

\begin{figure}[tbp]
\centering % \begin{center}/\end{center} takes some additional vertical space
\includegraphics[width=.7\textwidth]{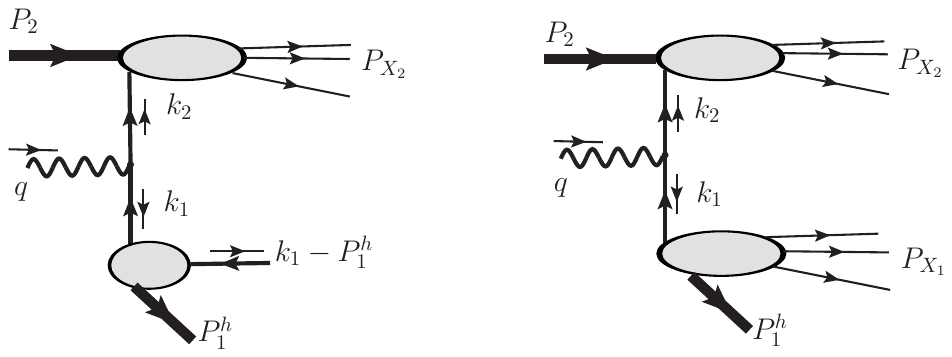}
% "\includegraphics" is very powerful; the graphicx package is already loaded
\vspace{-10cm}
\caption{\label{Fig-S-1} The hadron production in the semi-exclusive process ($K=k_1-P_1^h$), see the left panel, and
in the semi-inclusive process, see the right panel.}
\end{figure}

The Sudakov decompositions take the standard forms
(for the sake of shortness, we omit the four-dimension indices)
\cite{Anikin:2017azc, Anikin:2017pch, Anikin:2017vvn, Anikin:2018fom}
\begin{eqnarray}
\label{Sudakov-decom-1}
&&P_2\approx \frac{\tilde Q}{y_B \sqrt{2}}\, n  + P_{2}^{\perp},\quad
P_1^h\approx \frac{\tilde Q}{x_B \sqrt{2}}\, n^*  + P_{1}^{h\,\perp} \, ,
\\
&&
n^*=(1/\sqrt{2},\,{\bf 0}_T,\,1/\sqrt{2}), \quad n=(1/\sqrt{2},\,{\bf 0}_T,\,-1/\sqrt{2})
\nonumber\\
&&S\approx \frac{\lambda}{M_2}\,P_2 + S_\perp\,
\end{eqnarray}
for the hadron momenta and spin vector;
\begin{eqnarray}
\label{Sudakov-decom-2}
q=\frac{\tilde Q}{\sqrt{2}}\, n^* + \frac{\tilde Q}{\sqrt{2}}\,n + q_{\perp}, \quad q_\perp  \ll
\tilde Q = \sqrt{-Q^2},
\end{eqnarray}
for the photon momentum.
The hadron momenta $P_1^h$ and $P_2$ have the plus and minus dominant light-cone
components, respectively, see Fig.~\ref{Fig-S-1}.

The differential cross-section takes therefore the following form
\begin{eqnarray}
\label{x-s-1}
&& d\sigma = \frac{d^3 \vec{{\bf l}}_2}{(2\pi)^3 2 l_2^0} \frac{d^3 \vec{{\bf P}}^h_1}{(2\pi)^3 2 E^h}
{\cal L}_{\mu\nu} {\cal W}_{\mu\nu},
\end{eqnarray}
where ${\cal L}_{\mu\nu}$ and ${\cal W}_{\mu\nu}$  denote the lepton and hadron tensors, respectively.

%
%
%%%%%%%%%%%%%%%%%%%%%%%%%%%%%%% FIGURE %%%%%%%%%%%%%%%%%%%%%%%%%%%%%%%%
%%%%%%%%%%%%%%%%%%%%%%%%%%%%%%%%%%%%%%%%%%%%%%%%%%%%%%%%%%%%%%%%%%%%%%%
%
%%%%%%%%%

%%%%%%%%%%%%%%%%%%%%%%%%%%%%%%%%%%%%%%%%%
\section{The hadron production in semi-exclusive processes}
\label{Sec:HT-DY}
%%%%%%%%%%%%%%%%%%%%%%%%%%%%%%%%%%%%%%%%%

The process under our discussion resembles the standard Drell-Yan process, see Fig.~\ref{Fig-S-1},
where one of the initial hadron states is being replaced by the final hadron state.
Moreover, all unobserved hadron states $|P_X \rangle$
can be divided into two (after the applied factorization procedure, independent)
sets, {\it i.e.} $|P_X \rangle = |P_{X_1} \rangle \oplus |P_{X_2} \rangle$.
The set given by $|P_{X_2}\rangle$ is associated with the upper part of the hadron tensor,
while the other set given by $| P^h_1, P_{X_1} \rangle = | P^h_1, K \rangle + ...$ (see Fig.~\ref{Fig-S-1})
is forming the lower part of hadron tensor.
In this context, we mainly follow the stages of our analysis that have been used in  a series of papers
\cite{Anikin:2017azc, Anikin:2017pch, Anikin:2017vvn, Anikin:2018fom}.
So, by definition, the hadron tensor of (\ref{proc-1}) is defined to be given by the following expression
\begin{eqnarray}
\label{h-t-1}
&&{\cal W}_{\mu\nu} = \int (d{\mathbb P}{\mathbb S}) \delta^{(4)}(
q + P_2
- P_1^h - K - P_{X_2}
)
\times
\nonumber\\
&&
\langle P_2 | J^{em}_{\nu}(0) | P_1^h, K, P_{X_2} \rangle
\langle P_1^h, K, P_{X_2}| J^{em}_{\mu}(0) | P_2 \rangle,
\end{eqnarray}
where the matrix elements are written in the Heisenberg representation and
the phase space measure is defined by
\begin{eqnarray}
\label{PS-1}
(d{\mathbb P}{\mathbb S}) =
\frac{d^3 \vec{{\bf K}}}{(2\pi)^3 2 K_0}
\frac{d^3 \vec{{\bf P}}_{X_2}}{(2\pi)^3 2 E_{X_2}}.
\end{eqnarray}
Notice that from the viewpoint of factorization theorem
the hadron tensor of (\ref{h-t-1}) is not yet in a factorized form.
To apply the factorization procedure, the hadron momenta $P_2$ and
$P^h_1$ should be associated with the light-cone minus and plus directions,
respectively, which are well-separated. In other words, there are no the strong interferences between the
hadron dominant directions.

In addition, it is important to stress that
due to the used factorization procedure the above-mentioned two sets of states,
$|P_{X_2}\rangle$ and  $| P^h_1, P_{X_1} \rangle \Rightarrow | P^h_1, K \rangle$,
are independent ones
because the light-cone plus direction, dominated in the lower blob, and light-cone minus direction,
dominated in the upper blob, are separated well.

With these, the Fock states corresponding to the initial and final hadrons
can be permuted resulting in
\begin{eqnarray}
\label{h-t-2}
{\cal W}_{\mu\nu} \sim
&&\Big\{ \langle 0|  \bar\psi(0)  | P_1^h, K \rangle \, \gamma_\nu \,\langle P_2 | \psi(0) |P_{X_2} \rangle \Big\}
\times
\nonumber\\
&&
\Big\{ \langle P_{X_2}| \bar\psi(0) | P_2 \rangle  \,\gamma_\mu \, \langle P_1^h, K | \psi(0) | 0 \rangle \Big\}
\end{eqnarray}
for the matrix elements in (\ref{h-t-1}).
Here, as mentioned, the unobserved hadrons
with $P_{X_2}$ are fully associated with the upper part of the process,
see Fig.~\ref{Fig-S-1} (the left panel),
and they are independent from the detected hadron with $P^h_1$.

In what follows, since the states with $P_{X_1}$ have been traded for the (jet) quark with $K$  and
the only-remained unobserved states are the states with $P_{X_2}$, we simplify the notations
 by the tacit replacement $P_{X_2} \equiv P_{X}$ unless it leads to some misunderstanding.

%%%%%%%%%%%%%%%%%%%%%%%%%%%
\subsection{The hadron tensor at leading order}
\label{SubSec:HT-LO}
%%%%%%%%%%%%%%%%%%%%%%%%%%%

In the paper, our main interest is associated with the gluon pole generated by the radiative corrections.
Nevertheless, we begin our consideration with the detail analysis of leading order hadron tensor.

Making used the Fourier transform for the delta-function  and the translation transforms in Eqn.~(\ref{h-t-1}),
the leading order hadron tensor ${\cal W}_{\mu\nu}^{(0)}$ can readily be written down as
\begin{eqnarray}
\label{h-t-3}
{\cal W}_{\mu\nu}^{(0)} = \int (d^4 \xi) e^{iq\xi}
{\rm tr} \big[ \gamma_\nu \, \Phi(\xi) \,\gamma_\mu \, \bar\Phi(\xi) \big],
\end{eqnarray}
where the hidden spinor indices related to the functions $\Phi$ and $\bar\Phi$ are open,
and
\begin{eqnarray}
\label{Phi-Func-1}
&&\hspace{-0.5cm}\Phi(\xi) = \sum_{X}\int \frac{d^3 \vec{{\bf P}}_X}{(2\pi)^3 2 E_X}
\langle P_2 | \psi(\xi) |P_X \rangle \, \langle P_X| \bar\psi(0) | P_2 \rangle,
\\
\label{Phi-Func-2}
&&\hspace{-0.5cm}\bar\Phi(\xi) = \int\frac{d^3 \vec{{\bf K}}}{(2\pi)^3 2 K_0}
\langle 0|  \bar\psi(\xi)  | P_1^h, K \rangle \, \langle P_1^h, K | \psi(0) | 0 \rangle.
\end{eqnarray}
In (\ref{h-t-3}), the functions $\Phi(\xi)$ and $\bar\Phi(\xi)$ can be moved apart to different positions in $x$-space
by the integration with the corresponding delta-functions, {\it i.e.}
\begin{eqnarray}
\label{h-t-4}
&&{\cal W}_{\mu\nu}^{(0)}= \int (d^4 \xi) e^{iq\xi} \int (d^4\eta_1) (d^4 \eta_2)
\times
\nonumber\\
&&
\delta^{(4)}(\eta_1 - \xi) \delta^{(4)}(\eta_2 - \xi)
{\rm tr} \big[ \gamma_\nu \, \Phi(\eta_1) \,\gamma_\mu \, \bar\Phi(\eta_2) \big].
\end{eqnarray}
Afterwards, a series of Fourier transforms gives the following representation for the hadron tensor
through the momentum loop integrations
\begin{eqnarray}
\label{h-t-5}
{\cal W}_{\mu\nu}^{(0)}&=& \int (d^4 k_1)  (d^4 k_2) \delta^{(4)}(k_1+k_2-q)
\times
\nonumber\\
&&
{\rm tr} \big[ \gamma_\nu \, \Phi(k_2) \,\gamma_\mu \, \bar\Phi(k_1) \big],
\end{eqnarray}
and
\begin{eqnarray}
\label{Phi-Func-1-2}
&&\hspace{-0.5cm}\Phi(k_2) = \int(d^4 \eta_2) e^{ik_2\eta_2} \times
\nonumber\\
&&
 \sum_{X}\int \frac{d^3 \vec{{\bf P}}_X}{(2\pi)^3 2 E_X}
\langle P_2 | \psi(\eta_2) |P_X \rangle \, \langle P_X| \bar\psi(0) | P_2 \rangle,
\\
\label{Phi-Func-2-2}
&&\hspace{-0.5cm}\bar\Phi(k_1) = \int (d^4 \eta_1) e^{ik_1\eta_1} \times
\nonumber\\
&&
\int\frac{d^3 \vec{{\bf K}}}{(2\pi)^3 2 K_0}
\langle 0|  \bar\psi(\eta_1)  | P_1^h, K \rangle \, \langle K, P_1^h | \psi(0) | 0 \rangle.
\end{eqnarray}

These functions can be expressed in terms of the corresponding distribution functions after the use
of Lorentz decompositions.

%%%%%%%%%%%%%%%%%%%%%%%%%%%%%%%%%%%%
\subsection{The archetypal (original) factorization procedure}
\label{Subsec:Fact-Pr}
%%%%%%%%%%%%%%%%%%%%%%%%%%%%%%%%%%%%

The factorization theorem is the main
tool of the hadron tensor calculation.
Generally speaking, the QCD factorization theorem, as one of asymptotic methods, allows
only the estimation of the amplitudes or hadron tensors instead of exact calculations due to
the quark confinement problem.

It is worth to remind that the original factorization theorem states the following:
if the kinematics has a large parameter, the short (hard) and long (soft) distance
dynamics can be independently separated out
during the asymptotical estimation procedure applied to given amplitudes (or hadron tensors).
{\it The final result of such estimation or factorization has to be presented in the form of
the mathematical convolution of the hard and soft parts, which are independent each other.
The mathematical convolution implies that the products of hard and soft parts are integrated out over the dimensionless
parton fractions.
}
\cite{Efremov:1976ih,Efremov:1978rn,Efremov:1978xm,Efremov:1978cu}.

To illustrate this typical factorization procedure, we consider
the arbitrary DY-like hadron tensor which involves two non-perturbative blobs, it reads
(here, for the sake of brevity, we omit all possible Lorentz indices)
\begin{eqnarray}
\label{Fact-1}
W=\int (d^4 k_1) (d^4 k_2) E(k_1, k_2, q) \Phi_1(k_1) \bar\Phi_2(k_2),
\end{eqnarray}
where
\begin{eqnarray}
\label{E-Phi}
&&E(k_1, k_2, q) = \delta^{(4)}(k_1+k_2-q) \, {\cal E}(k_1, k_2, q)
\nonumber\\
&&
\Phi_1(k_1)\stackrel{{\cal F}_1}{=} \langle \bar\psi(z_1) \Gamma_1 \psi (0) \rangle,
\nonumber\\
&&
\bar\Phi_2(k_2)\stackrel{{\cal F}_2}{=} \langle \bar\psi(0) \Gamma_2 \psi (z_2) \rangle
\end{eqnarray}
and $\stackrel{{\cal F}_i}{=}$ denotes the corresponding Fourier transforms.
In Eqn.~(\ref{E-Phi}), ${\cal E}(k_1, k_2, q)$ indicates the product of different propagators
appearing after the use of Wick's theorem, while the $\delta$-function reflects the corresponding
momentum conservation at the subprocess level, see Eqns.~(\ref{h-t-3})-(\ref{h-t-5}).

We emphasize that the representation of hadron tensor given by Eqn.~(\ref{Fact-1})
is exact, before the factorization procedure has been applied. One of the features pointing out that we deal with
the non-factorized representation is the presence of dimensionful loop integrations
binding the product of propagators with the non-perturbative functions.
The dimensionful loop integrations do {\it not} allow to treat the product of propagators,
which pretends to form the hard part, as independent of the non-perturbative (soft) functions.

We begin with choosing the dominant directions dictated by the given process kinematics.
In the example of Eqn.~(\ref{Fact-1}), we have two dominant directions:
one of them is associated with the plus light-cone direction, the other to the minus light-cone
direction. The important condition for factorization is that the dominant directions must be
separated rather well in order to be independent ones.

In the next step, we have to introduce the definitions of the dimensionless parton fractions with the help of
the following replacement:
\begin{eqnarray}
\label{Fth-Rep}
d^4 k_i \Rightarrow d^4 k_i \int_{-1}^{+1}dx_i \delta(x_i - k_i^{\pm}/P_i^{\pm})
\end{eqnarray}
depending on the chosen dominant direction.
The physical spectral properties of fractions
are determined by the $\alpha$-representation where the estimation procedure should be established.

Further, we expand the product of propagators (together with the $\delta$-function, see Eqn.~(\ref{delta-decom}))
around the chosen dominant directions.
As a result, we obtain that
\begin{eqnarray}
\label{Fact-2}
&&W^{(0)}=\int (d x_1) (d x_2) E(x_1P^+_1, x_2P^-_2; q)
\nonumber\\
&&
\times
\Big\{
\int (d^4 k_1) \delta(x_1 - k_1^{+}/P_1^{+})
\Phi_1(k_1)
\Big\}
\nonumber\\
&&
\times
\Big\{
\int (d^4 k_2)  \delta(x_2 - k_2^{-}/P_2^{-})
\bar\Phi_2(k_2)
\Big\}
\end{eqnarray}
 if we neglect the $k^\perp_i$-terms in the expansion; and
 \begin{eqnarray}
\label{Fact-3}
&&W^{(i,j)}=\int (d x_1) (d x_2) \sum_{i,j}E^{(i,j)}(x_1P^+_1, x_2P^-_2; q)
\nonumber\\
&&
\times
\Big\{
\int (d^4 k_1) \delta(x_1 - k_1^{+}/P_1^{+}) \,
\prod_{i^\prime=1}^{i}
k_{1\, i^\prime}^\perp\,
\Phi_1(k_1)
\Big\}
\nonumber\\
&&
\times
\Big\{
\int (d^4 k_2)  \delta(x_2 - k_2^{-}/P_2^{-})\,
\prod_{j^\prime=1}^{j}
 k_{2\, j^\prime}^\perp\,
\bar\Phi_2(k_2)
\Big\}
\end{eqnarray}
if the transverse momentum terms are essential in the expansion.
Notice that in Eqn.~(\ref{Fact-3}) the $k_i^\perp$-dependence appears in the form of
the corresponding integral moments only. Moreover, the parts of $W^{(0)}$
and $W^{(i,j)}$ with $E$-function and $\Phi$-functions are not
related by the integration over $k_i^\perp$.

Last but not least, in order to achieve the results of Eqns.~(\ref{Fact-2}) and (\ref{Fact-3}),
in the DY-like processes the four-dimensional $\delta$-function with the argument describing
the momentum conservation at the parton level
should be treated as the ``hard'' parts of hadron tensor.
This statement has been proven with the help of the factorization links introduced in \cite{Anikin:2006hm}.
In particular,
the decomposition of delta function around the hadron dominant directions takes the form of
\begin{eqnarray}
\label{delta-decom}
&&\delta^{(4)}(k_1+k_2-q) = \delta^{(4)}(x_1P_1+x_2P_2-q) +
\nonumber\\
&&
\frac{\partial \delta^{(4)}(k_1+k_2-q)}{\partial k_i^\rho} \Big |_{k_i=x_iP_i}\biggr. \, k_{i\, \perp}^\rho + \ldots
\end{eqnarray}
Notice that for our purposes it is enough to be limited by the first term of decomposition in the {\it r.h.s.} of (\ref{delta-decom}).
The other important feature of $\delta^{(4)}(\langle mom. conserv. \rangle)$ is that
the momentum conservation law is valid for any system chosen for the kinematical reason.
In this context, we may consider this delta function as a kind of the Lorentz invariant despite the covariant form of momentum conservation.

Thus, once the factorization procedure applied, the leading order hadron tensor takes the factorized form
\begin{eqnarray}
\label{h-t-4-1}
{\cal W}_{\mu\nu}^{(0)}= &&\delta^{(2)}(\vec{\bf q}_\perp)
\int (d x)  (d y) \delta(x P^{h\, +}_1 - q^+)
\delta(yP^{-}_2 -q^-)
\nonumber\\
&&
\times
{\rm tr} \big[ \gamma_\nu \, \Phi(y) \,\gamma_\mu \, \bar\Phi(x) \big]\,,
\end{eqnarray}
where, as stressed above, the mathematical convolution is given by the integration
over dimensionless fractions $x$ and $y$.
In other words, in our case, the parts of ${\cal W}_{\mu\nu}^{(0)}$
with the delta function and $\bar\Phi$- and $\Phi$-functions are not linked by the integration over
the dimensionful $k_i^\perp$.

There are, however, the alternative
approaches which are based on the formalism with the nonzero photon transverse momentum
and without the $\delta$-function expansion,  see \cite{Barone:2001sp}. These approaches result in
the leading order hadron tensor with the $k^\perp$-dependent non-perturbative functions
which has the following form
\begin{eqnarray}
\label{NoFact-1}
&&\tilde W^{(0)}=\int (d x_1) (d x_2) {\cal E}(x_1P^+_1, x_2P^-_2; q)
\nonumber\\
&&
\times
\Big\{
\int (d^2 \vec{\bf k}^\perp_1)
(d^2 \vec{\bf k}^\perp_2)
\delta^{(2)} (\vec{\bf k}^\perp_1 + \vec{\bf k}^\perp_2 - \vec{\bf q}^\perp)
\nonumber\\
&&
\times
\int (dk^+_1 d k_1^-) \delta(x_1 - k_1^{+}/P_1^{+})
\Phi_1(k_1)
\nonumber\\
&&
\times
\int (d k^-_2 dk^+_2)  \delta(x_2 - k_2^{-}/P_2^{-})
\bar\Phi_2(k_2)
\Big\}\, e^{-S(k^\perp_1/\Lambda,\,  k^\perp_2/\Lambda) - (\text{the other sources of} \,\, k_\perp)}.
\end{eqnarray}
Here, generally speaking,
the functions $\Phi(k_1)$ and $\bar\Phi(k_2)$ cannot be considered as the independent of each other
because the two-dimensional integrations with the $\delta$-function relate them.
This would lead to the factorization breaking effects.
However, the additional
Sudakov-like exponentials of $e^{-S(\vec{\bf k}^2_\perp/\Lambda^2)}$-form,
generated by the corresponding summation of
gluon radiations, should minimize the mentioned non-factorized effects.
This way of proceeding is really complicated and it demands a very careful analysis which is beyond of our paper.

%%%%%%%%%%%%%%%%%%%%%%%%%%%%%%%%%%%
\subsection{The leading order function $\bar\Phi(k_1)$}
\label{SubSec:LO-Function}
%%%%%%%%%%%%%%%%%%%%%%%%%%%%%%%%%%%

\begin{figure}[tbp]
\centering % \begin{center}/\end{center} takes some additional vertical space
\includegraphics[width=.48\textwidth]{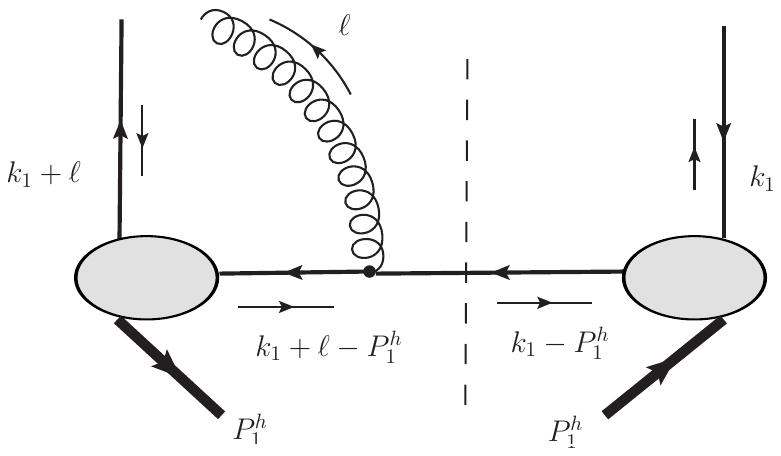}
\,
\includegraphics[width=.48\textwidth]{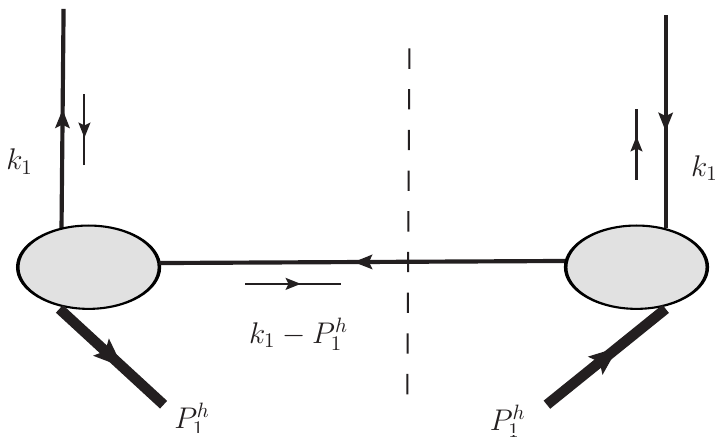}
% "\includegraphics" is very powerful; the graphicx package is already loaded
\vspace{-6cm}
\caption{\label{Fig-S-2} The hadron production in the semi-exclusive process: the next-to-leading function $\bar\Phi^{(A)}(k_1, \ell)$
(the left panel) and the leading order function $\bar\Phi(k_1)$ (the right panel).}
\end{figure}

Let us now dwell on the function $\bar\Phi(k_1)$ of (\ref{Phi-Func-2-2}), see Fig.~\ref{Fig-S-2}, the right panel.
This function contains the integration over the intermediate on-shell (anti)quark Fock state
\begin{eqnarray}
\label{Inter-st-1}
&&{\cal I}_K=\int\frac{d^3 \vec{{\bf K}}}{(2\pi)^3 2 K_0}
| K \rangle \, \langle  K | =
\nonumber\\
&&
\int\frac{d^3 \vec{{\bf K}}}{(2\pi)^3 2 K_0} d^+_\lambda(K) | 0\rangle \, \langle 0 | d_\lambda(K)
\end{eqnarray}
which can also be presented as
\begin{eqnarray}
\label{Inter-st-2}
&&{\cal I}_K=
\frac{E}{m_q}
\int (d^4 P) \delta(P^2) \int (d^4 K) \delta(K^2)
\delta^{(3)} (\vec{\bf P} - \vec{\bf K}) \times
\nonumber\\
&&
[d^+_\lambda(K) v_\lambda(K)] [d_{\lambda^\prime}(P) \bar v_{\lambda^\prime}(P)] \big |_{K_0=P_0=E},
\end{eqnarray}
where $E$ and $m_q$ denote the energy and mass of the intermediate (anti)quark, respectively.

In (\ref{Inter-st-2}), $m_q$ is a result of the normalization condition
\begin{eqnarray}
\label{Norm-Cond}
v_\lambda(K) \bar v_{\lambda^\prime}(K)=2m_q \delta_{\lambda\lambda^\prime},
\end{eqnarray}
and the integral unit given by
\begin{eqnarray}
\label{Int-Unit-1}
&&{\bf 1} = \int (d^3 \vec{\bf P}) \, \delta^{(3)} (\vec{\bf P} - \vec{\bf K})
\nonumber\\
&&
=
2E \int (d^4 P) \, \delta( P^2 ) \, \delta^{(3)} (\vec{\bf P} - \vec{\bf K})
\end{eqnarray}
has been used.
Also, the imposed condition $K_0=P_0=E$, which is shown in Eqn.~(\ref{Inter-st-2}),
can be written in the equivalent form as
$E \delta\left( P_0 - K_0\right)$.
All these tricks lead to the following representation
\begin{eqnarray}
\label{Inter-st-3}
&&{\cal I}_K=
\frac{E^2}{m_q}
\int (d^4 P) \int (d^4 K)
\delta^{(4)} (P -  K) \times
\\
&&
\Big\{ \int (d^4 x) e^{-iKx} \psi_{\text{on-sh.}}(x)\Big\} \Big\{ \int (d^4 y) e^{iPy} \bar\psi_{\text{on-sh.}}(y)\Big\}.
\nonumber
\end{eqnarray}
Here, $\psi_{\text{on-sh.}}$ denotes the on-shell particle (similary, for $\bar\psi_{\text{on-sh.}}$), {\it i.e.}
\begin{eqnarray}
\label{q-on}
\int (d^4 x) e^{-iKx} \psi_{\text{on-sh.}}(x)=\delta(K^2) [d^+_\lambda(K) v_\lambda(K)].
\end{eqnarray}
Hence, the function $\bar\Phi(k_1)$ of Eqn.~(\ref{Phi-Func-2-2}) takes the form of
\begin{eqnarray}
\label{Phi-Func-2-3}
\bar\Phi(k_1) = \frac{E^2}{m_q}\, \bar\Phi^{(2)}_{\text{o}}(P^h_1-k_1)\,
\bar\Phi^{(1)}_{\text{o}}(k_1)
\end{eqnarray}
where
\begin{eqnarray}
\label{Phi-Func-3-1}
\bar\Phi^{(1)}_{\text{o}}(k_1)=
\int (d^4 \xi_1) e^{ik_1\xi_1}
\langle 0 |  \bar\psi(\xi_1)  \psi_{\text{on-sh.}}(0) | P_1^h \rangle
\end{eqnarray}
and
\begin{eqnarray}
\label{Phi-Func-3-2}
&& \bar\Phi^{(2)}_{\text{o}}(P^h_1-k_1)=
\nonumber\\
&&
\int (d^4 \eta_1) e^{-i(P^h_1-k_1)\eta_1}
\langle P_1^h |  \bar\psi_{\text{on-sh.}}(\eta_1)  \psi(0) | 0 \rangle.
\end{eqnarray}
For the sake of completeness, it is instructive to give the representation of $\bar\Phi(k_1)$ in Eqn.~(\ref{Phi-Func-2-3})
if the factorization procedure has been applied. It reads (here, $\bar x_1 = 1- x_1$)
\begin{eqnarray}
\label{Phi-Func-2-4}
&&\bar\Phi(x_1) =  \int (d^4 k_1) \delta(x_1 - k_1^+/P_1^{h\, +}) \, \bar\Phi(k_1)=
\nonumber\\
&&
P_1^{h, +}  \int (d k_1^-)(d^2 \vec{\bf k}_{1}^{\perp})
\nonumber\\
&&
\times
\frac{E^2}{m_q}
\bar\Phi^{(2)}_{\text{o}}(\bar x_1, k_1^-, \vec{\bf k}_{1}^{\perp})\,
\bar\Phi^{(1)}_{\text{o}}(x_1,  k_1^-, \vec{\bf k}_{1}^{\perp}).
\end{eqnarray}
It is worth to notice that there is an alternative way to present the functions like $\bar\Phi(k_1)$
involving the on-shell fields in correlators, see App.~\ref{AppA} for details.

As above mentioned, the intermediate quark (see, for example, Eqn.~(\ref{Phi-Func-2-4})) is an on-shell particle,
{\it i.e.}
\begin{eqnarray}
\label{on-q}
&&(k_1-P^h_1)^2\equiv (k_1^+ - P^{h\,+}_1) k_1^- - (\vec{\bf k}^\perp_1-\vec{\bf P}^{h\,\perp}_1)^2 = m^2_q
\nonumber\\
&&
\Longrightarrow (k_1^+ - P^{h\,+}_1) =\frac{m^2_q}{k^-_1},
\end{eqnarray}
where the condition $|\vec{\bf k}^\perp_1| \sim |\vec{\bf P}^{h\,\perp}_1| \ll P^{h\,+}_1$ has been used.
Therefore, using $k_1^+ = x P^{h\,+}_1$, Eqn.~(\ref{on-q}) shows that the on-shell quark condition
corresponds to the limits of $x\to 1$ and $m_q\to 0$.

%
%
%%%%%%%%%%%%%%%%%%%%%%%%%%%%%%% FIGURE %%%%%%%%%%%%%%%%%%%%%%%%%%%%%%%%
%%%%%%%%%%%%%%%%%%%%%%%%%%%%%%%%%%%%%%%%%%%%%%%%%%%%%%%%%%%%%%%%%%%%%%%
%
%%%%%%%%%

%%%%%%%%%%%%%%%%%%%%%%%%%%%%%%%%%%%%%%
\subsection{Hadron tensor with the radiative correction before factorization}
\label{SubSec:HT-NLO-1}
%%%%%%%%%%%%%%%%%%%%%%%%%%%%%%%%%%%%%%

\begin{figure*}[ht]
\begin{center}
\includegraphics[width=0.48\textwidth]{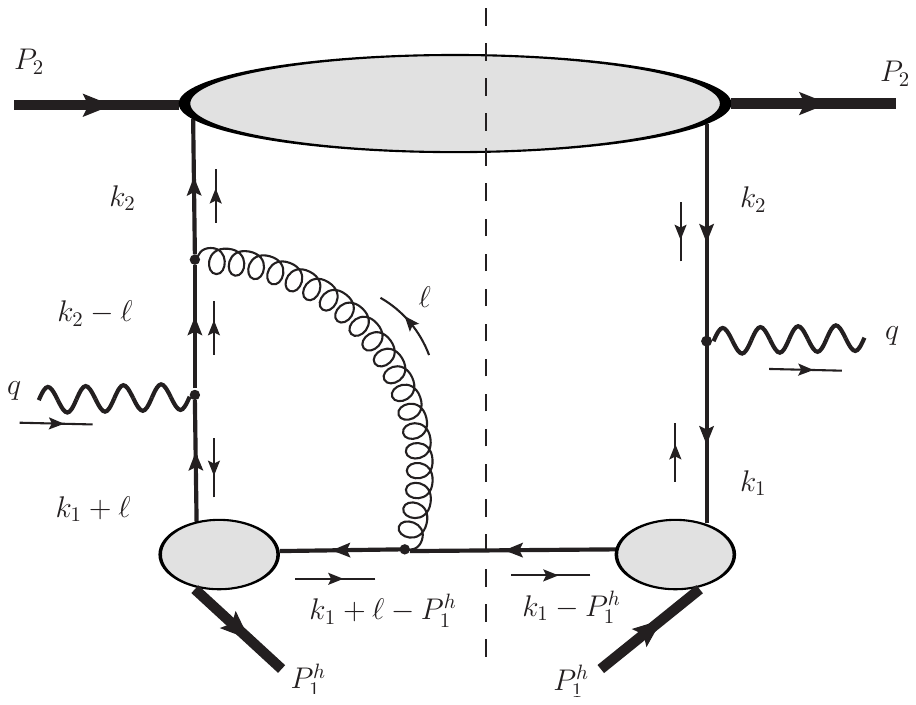}
\,
\includegraphics[width=0.48\textwidth]{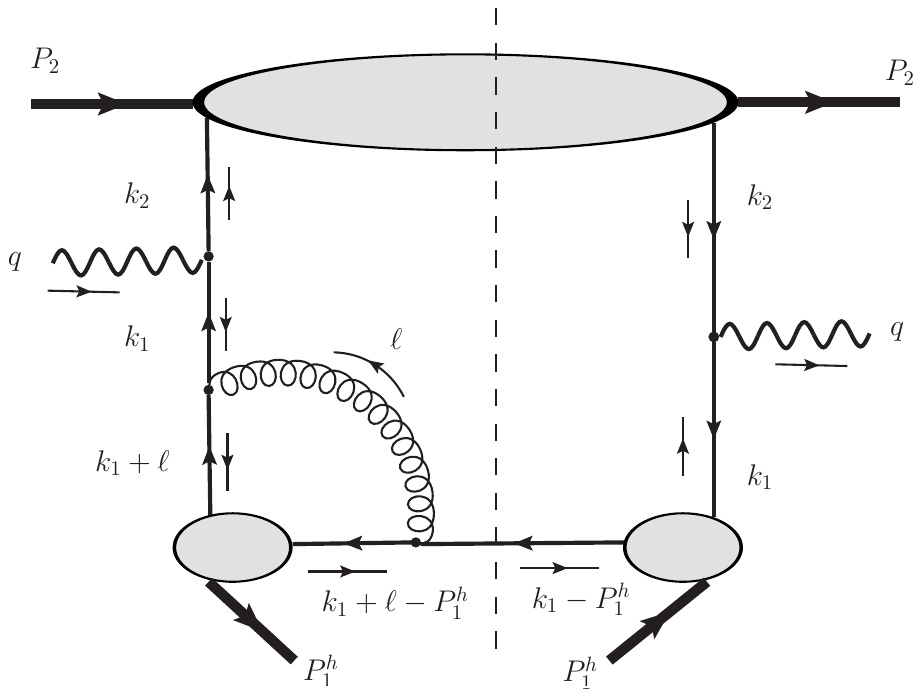}
\end{center}
\vspace{-5.5cm}
\caption{The hadron production in the semi-exclusive process: the standard diagram contribution
(the left panel) and the non-standard diagram contribution (the right panel)
to the gauge invariant hadron tensor. The dominant quark and gluon momenta $k_1$ and $\ell$ lie
along the plus direction while the dominant antiquark momentum $k_2$ -- along the minus direction.}
\label{Fig-S-3}
\end{figure*}

A straightforward extension of \cite{Anikin:2017azc, Anikin:2017pch, Anikin:2017vvn, Anikin:2018fom}
gives the sum of standard and non-standard contributions to the hadron tensor
with radiative corrections, see Fig.~\ref{Fig-S-2}, the left panel, and Fig.~\ref{Fig-S-3}.

Before the factorization procedure applied, we have
\begin{eqnarray}
\label{HadTen-St}
&&{\cal W}_{\mu\nu}^{(\text{stand.})}=\int (d^4k_1)\,(d^4k_2) \, \delta^{(4)}(k_1+k_2-q)
\nonumber\\
&&\times\int (d^4\ell)\, {\cal D}_{\alpha\beta}(\ell)
\text{tr}\big[ \gamma_\nu \Gamma \gamma_\alpha S(k_2-\ell) \gamma_\mu \Gamma_1 \gamma_\beta \Gamma_2
\big]
\nonumber\\
&&\times\Phi^{[\Gamma]}(k_2)\,
\bar\Phi_{(2)}^{[\Gamma_1]}(k_1; \ell)\,
\bar\Phi_{(1)}^{[\Gamma_2]}(k_1)\,
\tilde\delta\big( (P_1^h-k_1)^2 \big) \,,
\end{eqnarray}
for the standard diagram contribution;
while, the non-standard diagram contribution is given the following expression
\begin{eqnarray}
\label{HadTen-NonSt}
&&{\cal W}_{\mu\nu}^{(\text{nonstand.})}=\int (d^4k_1)\,(d^4k_2) \, \delta^{(4)}(k_1+k_2-q)
\nonumber\\
&&\times\int (d^4\ell)\, {\cal D}_{\alpha\beta}(\ell)
\text{tr}\big[ \gamma_\nu \Gamma \gamma_\mu S(k_1) \gamma_\alpha \Gamma_1 \gamma_\beta \Gamma_2
\big]
\nonumber\\
&&\times\Phi^{[\Gamma]}(k_2)\,
\bar\Phi_{(2)}^{[\Gamma_1]}(k_1; \ell)\,
\bar\Phi_{(1)}^{[\Gamma_2]}(k_1)\,
\tilde\delta\big( (P_1^h-k_1)^2 \big).
\end{eqnarray}
The non-perturbative functions of Eqns.~(\ref{HadTen-St}) and (\ref{HadTen-NonSt}) are defined as
\begin{eqnarray}
\label{barPhi-func}
&&\Phi^{[\Gamma]}(k_2)= \int\hspace{-0.4cm} \sum\nolimits_{X}\,\int (d^4\eta_2)\, e^{-ik_2\eta_2}\,
\\
&&\times\langle P_2, S_\perp |\text{tr}\big[ \psi(0)|P_{X}\rangle \, \langle P_{X} | \bar\psi(\eta_2) \Gamma \big] | S_\perp , P_2\rangle\
\nonumber\\
&&
\equiv
- \int (d^4\eta_2)\, e^{-ik_2\eta_2}\,
\langle P_2, S_\perp | \bar\psi(\eta_2) \Gamma \psi(0) | S_\perp , P_2\rangle\
\nonumber
\end{eqnarray}
and
\begin{eqnarray}
\label{Phi2-func}
&&\bar\Phi_{(2)}^{[\Gamma_1]}(k_1; \ell)=
\\
&& \int (d^4\eta_1)\, e^{- i(P_1^h-\ell-k_1)\eta_1}\, \langle P_1^h | \bar\psi(\eta_1) \,\Gamma_1 \psi(0) | 0 \rangle\,,
\nonumber
\end{eqnarray}
\begin{eqnarray}
\label{Phi1-func}
&&\bar\Phi_{(1)}^{[\Gamma_2]}(k_1)= \int (d^4\xi)\, e^{ik_1\xi}\,
\langle 0 | \bar\psi(\xi) \,\Gamma_2 \psi(0) | P_1^h \rangle\,,
\end{eqnarray}
where the correlators contain the off-shell fermion operators, cf. Eqns.~(\ref{Phi-Func-3-1}) and (\ref{Phi-Func-3-2}).
Moreover,
for the sake of practical application, we introduce the dimensionless analog of delta-function
in the form of (see appendix~\ref{AppA})
\begin{eqnarray}
\label{dim-less-delta}
&&\tilde\delta\big( (P_1^h-k_1)^2 \big) =
\\
&&
\lim_{\{m_q, \,\bar x_1\}\to 0} \,
\frac{m_q^2}{2 \bar x_1 P_1^{h\, +}}
\delta\Big(
k_1^- - \frac{(\vec{\bf k}^\perp_1-\vec{\bf P}_1^{h\, \perp})^2+m_q^2}{2\bar x_1 P_1^{h\, +}}
\Big).
\nonumber
\end{eqnarray}
which ensures that one of fermions in the function
$\bar\Phi_{(1)}^{[\Gamma_2]}(k_1)$ becomes the on-shell one.

%%%%%%%%%%%%%%%%%%%%%%%%%%%%%%%%%%%%%%
\subsection{The next-to-leading order function $\bar\Phi^{(A)}(k_1, \ell)$
with gluon radiation}
\label{SubSec:NLO-Function}
%%%%%%%%%%%%%%%%%%%%%%%%%%%%%%%%%%%%%%

In the standard and non-standard hadron tensors, see Eqns.~(\ref{HadTen-St}) and (\ref{HadTen-NonSt}),
the next-to-leading function $\bar\Phi^{(A)}(k_1, \ell)$ (see Fig.~\ref{Fig-S-2}, the left panel)
extends the leading order function $\bar\Phi(k_1)$ (see Fig.~\ref{Fig-S-2}, the right panel) and
includes the functions $\bar\Phi_{(2)}(k_1; \ell)$
and $\bar\Phi_{(1)}(k_1)$ of Eqns.~(\ref{Phi2-func}) and (\ref{Phi1-func}) (here, we omit the $\Gamma$-structure in each functions).
The gluon momentum $\ell$ dependence of $\bar\Phi_{(2)}(k_1; \ell)$
indicates the fact that we deal with the next-to-leading order due to the gluon radiation.

To demonstrate this, we begin with the representation of the leading order function $\bar\Phi(k_1)$ given by Eqn.~(\ref{Phi-Func-2-3}).
We now take into account the first order of interaction in the correlator and write the following
\begin{eqnarray}
\label{Phi-Func-A-1}
&&\bar\Phi^{(A)}(k_1) = \frac{E^2}{m_q}\,
\Big\{
\int (d^4 \xi) e^{-i(P^h_1-k_1)\xi} \int (d^4 z)
\nonumber\\
&&
\times
\langle P_1^h |   \psi(0)
\big[ \bar\psi(z) \hat A(z) \psi(z)
\big]
\bar\psi_{\text{on-sh.}}(\xi)  | 0 \rangle
\Big\}
\nonumber\\
&&\times
\Big\{
\int (d^4 \eta) e^{ik_1\eta}
\langle 0 |  \psi_{\text{on-sh.}}(0)  \bar\psi(\eta) | P_1^h \rangle
\Big\}.
\end{eqnarray}
Here, the first figure-bracketed correlator with interactions has to be understood as the correlator
written in the interaction representation (the time-ordering symbol has been omitted).
Using Wick's theorem and Fourier transformations, we derive that
\begin{eqnarray}
\label{Phi-Func-A-2}
&&\bar\Phi^{(A)}(k_1) = \frac{E^2}{m_q}\,
\Big\{
 \int (d^4 z) e^{-i(P^h_1-k_1)z} \int (d^4 \ell)
\nonumber\\
&&
\times
\langle P_1^h |   \psi(0)
\bar S(k_1- P^h_1 + \ell) \hat A(\ell)
 \bar\psi_{\text{on-sh.}}(z)
 | 0 \rangle
\Big\}
\nonumber\\
&&
\times
\Big\{
\int (d^4 \eta) e^{ik_1\eta}
\langle 0 |  \psi_{\text{on-sh.}}(0) \bar\psi(\eta)  | P_1^h \rangle
\Big\},
\end{eqnarray}
or in the equivalent form it reads
\begin{eqnarray}
\label{Phi-Func-A-3}
&&\bar\Phi^{(A)}(k_1) = \frac{E^2}{m_q}\,
\Big\{
 \int (d^4 \ell)\,
\bar\Phi^{(2)}_{\text{o}}(P^h_1-k_1)
\nonumber\\
&&\times
\bar S(k_1- P^h_1 + \ell) \hat A(\ell)
\Big\}\,
\bar\Phi^{(1)}_{\text{o}}(k_1).
\end{eqnarray}

%%%%%%%%%%%%%%%%%%%%%%%%%%%%%%%%%%%%%%%%%%%%%
\section{Dominant role of gluon pole contribution}
\label{Sec:Comp-LO-NLO}
%%%%%%%%%%%%%%%%%%%%%%%%%%%%%%%%%%%%%%%%%%%%%

In this section, we give a proof that
the next-to-leading order (gluon pole contribution)  function
$\bar\Phi^{(A)}(k_1, \ell)$ is dominating over
the leading order function $\bar\Phi(k_1)$  in the kinematical region of $x_1\to 1$ and $m_q\to 0$.
We remind that the limits of $x_1\to 1$ and $m_q\to 0$ correspond to the fact that the intermediate
quark is on-shell, see Eqn.~(\ref{on-q}).
The result of this section can be considered as a principle finding of our study.

The kinematical and model-independent prefactor $E^2/m_q$ which appears in both the leading and next-to-leading
functions plays the crucial role in our analysis.
Namely, one can see that $E^2/m_q$ at  the leading function $\bar\Phi(k_1)$ of Eqn.~(\ref{Phi-Func-2-3})
nullify this contribution in the certain domain. Indeed,
using the kinematical constrains inspired by (\ref{Sudakov-decom-1}), we have
\begin{eqnarray}
\label{E-lc}
&&\bar\Phi(k_1) \Rightarrow
\frac{E^2}{m_q} = \frac{(k_1^+ - P_1^{h, +} + k_1^-)^2}{2 m_q}
\\
&&
\Rightarrow \frac{(\bar x_1 P_1^{h, +} + k_1^-)^2}{2 m_q} \Big |^{x_1\to1}_{ m_q\to 0}
\sim
\frac{\lim_{\epsilon\to 0} \epsilon^2}{\lim_{\delta\to 0} \delta} \sim
\lim_{\epsilon\to 0} \epsilon \equiv [0]^1,
\nonumber
\end{eqnarray}
where $\sim$ means ``behaves as''.

On the other hand, the situation is changed drastically if we consider the
the next-to-leading function related to the radiative correction. Focusing on
the gluon pole contribution to Eqn.~(\ref{Phi-Func-A-3}), which exists at $x_2=x_1$,
we obtain that
\begin{eqnarray}
\label{Behav-limit}
&&
\bar\Phi^{(A)}(k_1) \Big |^{x_1\to1}_{ m_q\to 0} \Rightarrow
\nonumber\\
&&
\frac{(k_1^+ - P_1^{h, +} + k_1^-)^2}{2 m_q} \bar S(k_1- P^h_1 + \ell) \Big |^{x_1\to1}_{ m_q\to 0}=
\frac{(k_1^+ - P_1^{h, +} + k_1^-)^2}{2 m_q} \frac{\gamma^+}{k^+_1- P^{h\,+}_1 + \ell^+} \Big |^{x_1\to1}_{ m_q\to 0}
\nonumber\\
&&
\sim \frac{\lim_{\epsilon\to 0} \epsilon^2}{\lim_{\delta\to 0} \delta \, \lim_{\eta\to 0} \eta}
\sim
\lim_{\epsilon\to 0} \frac{\epsilon^2}{\epsilon^2} \equiv [1],
\end{eqnarray}
where the representation given by
\begin{eqnarray}
\label{Prop-1}
k^+_1- P^{h\,+}_1 + \ell^+ = -\bar x_1 P^{h\,+}_1 + (x_2-x_1) P^{h\,+}_1
\end{eqnarray}
has been applied. For the function $\bar\Phi^{(A)}(k_1)$, we are interested in the combination of
$\bar\Phi^{(2)\, \text{tw-2}}_{\text{o}} \otimes \bar\Phi^{(1)\, \text{tw-3}}_{\text{o}}$ which results in
the $\gamma^+$-term of the propagator in the second line of
Eqn.~(\ref{Behav-limit}).
Thanks for Eqn.~(\ref{Behav-limit}), one can see that $[0]$ stemming from the combination $E^2/m_q$
is compensated by $[0]$ from the (anti)quark propagator if we deal with the gluon pole contribution.

It is important to stress that an alternative consideration of function $\bar\Phi^{(A)}(k_1)$ described in Appendix~\ref{AppA}
leads to the same behaviour of  $\bar\Phi^{(A)}(k_1)$ as in Eqn.~(\ref{Behav-limit}).

Therefore, we can conclude that the leading order hadron is actually suppressed in comparison with the
gluon pole contribution to the hadron tensor in the limits of  $x_1\to 1$ and $m_q\to 0$.
Based on that, we are able to concentrate entirely on the contributions associated
with the gluon poles.

%%%%%%%%%%%%%%%%%%%%%%%%%%%%%%%%%%%%%%%%%%%%%
\section{Gauge-invariant hadron tensor with the radiative correction in the
factorized form}
\label{Sec:GI-HT}
%%%%%%%%%%%%%%%%%%%%%%%%%%%%%%%%%%%%%%%%%%%%%

We are in a position to discuss the final expression for the hadron tensor.
We remind that
the concrete $\Gamma$ matrix projections (or the Fierz projections)  in the
corresponding functions depend on the considered case. For example,
in order to describe the pion or rho meson production, the corresponding functions receive the $\gamma$-structure
as
(see (\ref{HadTen-St}) and (\ref{Phi2-func}), (\ref{Phi1-func}))
\begin{eqnarray}
\label{gamma-str-N}
&&\Gamma\otimes \Phi^{[\Gamma]} \Rightarrow  \gamma^+
\otimes \Phi^{[\gamma^-]} \oplus \sigma^{+\perp}
\otimes \Phi^{[\sigma^{-\perp} ]}
\end{eqnarray}
for the leading twist two nucleon distribution functions;
\begin{eqnarray}
\label{gamma-str-M}
&&\Gamma_1\otimes \bar\Phi_{(2)}^{[\Gamma_1]} \Rightarrow  \gamma^-(\gamma_5) \otimes \bar\Phi_{(2)}^{[\gamma^+(\gamma_5)]},
\\
&&
\label{gamma-str-M-tw3}
\Gamma_2\otimes \bar\Phi_{(1)}^{[\Gamma_2]} \Rightarrow
\gamma_\rho^\perp (\gamma_5)
\otimes \bar\Phi_{(1)}^{[\gamma^\perp_\rho(\gamma_5)]}
 \oplus
 \sigma^{+ -} (\gamma_5)
 \otimes \bar\Phi_{(1)}^{[\sigma^{- +}(\gamma_5) ]}
\end{eqnarray}
for the pion and rho meson distribution amplitudes which correspond to the twist two (\ref{gamma-str-M}) and
the twist three (\ref{gamma-str-M-tw3}).

While the factorization procedure applied, we deal with the gauge invariant hadron tensor that takes the form of
\begin{eqnarray}
\label{HadTen-GI}
&&{\cal W}^{\mu\nu}=  \delta^{(2)}(\vec{\bf q}_\perp)
\int (dx_1)\,(dy) \, \delta(x_1P^{h,+}_1 - q^+)\delta(yP^-_2 - q^-)
\nonumber\\
&&\times  F(y) \int (dx_2) \widetilde{B}(x_1,x_2) \,\frac{T^\nu}{P^{h}_1\cdot P_2}
\Big[ \frac{P^{h\,\mu}_{1}}{y} - \frac{P^{\mu}_{2}}{x_1}\Big] \,,
\end{eqnarray}
where
\begin{eqnarray}
 F(y)=
          \begin{pmatrix}
           f_T(y)\vspace{0.2cm}\\
           h_1(y)
          \end{pmatrix},\quad
T^\nu =
          \begin{pmatrix}
          \varepsilon^{P^h_1+S_\perp P_2^\perp} V^\nu_{\perp}\vspace{0.2cm}\\
          \varepsilon^{\nu S_\perp P_2 P^h_1}_\perp
          \end{pmatrix}
\end{eqnarray}
with $V^\nu_{\perp}=\{ P^{h\,\perp}_{1}, \, e^\perp\}^\nu$.
The distribution function $F(y)$ represents the $k^\perp$-integrated function \cite{Barone:2001sp}.

The structure function (or distribution function) $\widetilde{B}(x_1,x_2)$ related to
gluon pole is defined as \cite{Anikin:2017azc}
\begin{eqnarray}
\label{B-fun}
\widetilde{B}(x_1,x_2)=\frac{1}{2}
\frac{\bar\Phi_{(1)}^{\text{tw-3}}(x_1)\hspace{-0.2cm}
\stackrel{\hspace{0.2cm}\vec{\bf k}_{1}^{\,\perp}}{\circledast}
\hspace{-0.1cm}\bar\Phi_{(2)}^{\text{tw-2}}(x_2)}{x_2-x_1 - i\epsilon},
\end{eqnarray}
where the transverse momentum integration is given by
\begin{eqnarray}
\label{Conv-Delta}
&&\bar\Phi_{(1)}^{\text{tw-3}}(x_1)\hspace{-0.2cm}
\stackrel{\hspace{0.2cm}\vec{\bf k}_{1}^{\,\perp}}{\circledast}
\hspace{-0.1cm}\bar\Phi_{(2)}^{\text{tw-2}}(x_2)=
\int(d^2\vec{\bf k}_{1}^{\,\perp})  \,
\bar\Phi_{(1)}^{\text{tw-3}}
(x_1; \vec{\bf k}_{1}^{\,\perp})\times
\nonumber\\
&&
\int(d\ell^- d^2\vec{\ell}_{\,\perp})
\frac{\bar\Phi_{(2)}^{\text{tw-2}}(\bar x_2; \vec{\bf k}_{1}^{\,\perp};\ell^- \vec{\ell}_{\,\perp})}
{\ell^- - \vec{\ell}^{\,2}_\perp/(2x_{21}P_1^{h\, +}) + i\,{\rm sign}(x_{21})\, \tilde\epsilon}.
\end{eqnarray}

%
%
%%%%%%%%%%%%%%%%%%%%%%%%%%%%%%% FIGURE %%%%%%%%%%%%%%%%%%%%%%%%%%%%%%%%
%%%%%%%%%%%%%%%%%%%%%%%%%%%%%%%%%%%%%%%%%%%%%%%%%%%%%%%%%%%%%%%%%%%%%%%
%
%%%%%%%%%

%%%%%%%%%%%%%%%%%%%%%%%%%%%%%%%%%%
\subsection{Hadron tensor with the radiative correction:
the case of $k_\perp$-dependent nucleon distribution functions}
\label{SubSec:HT-kT}
%%%%%%%%%%%%%%%%%%%%%%%%%%%%%%%%%%

The gauge invariant hadron tensor of (\ref{HadTen-GI}) can be readily extended to the case with the
essential $k_\perp$-dependence of the nucleon distribution function.
As the first stage, we consider the nucleon matrix element of the quark vector projection given by
(see (\ref{gamma-str-N}) and Appendix~\ref{NF-Par})
\begin{eqnarray}
\label{V-me}
\Phi^{[\gamma^-]}(k_2)=\,\varepsilon^{ -+ k^\perp_2 \alpha }
 \begin{pmatrix}
          f_{(1)} (y, \vec{\bf k}_2^\perp) \, S_{\alpha}^\perp \vspace{0.2cm}\\
          f_{(2)}(y, \vec{\bf k}_2^\perp)   \, s_{\alpha}^\perp \vspace{0.2cm}
          \end{pmatrix},
\end{eqnarray}
where $S_\alpha^\perp$ and $s_\alpha^\perp$ stand for the nucleon and quark spin covariant vectors, respectively.
Notice that the function $f_{(1)} (y, \vec{\bf k}_2^\perp)$ is nothing but the well-known function $f_{1\,T}^\perp (y, \vec{\bf k}_2^\perp)$,
while
the $k_\perp$-dependent function $f_{(2)}(y, \vec{\bf k}_2^\perp)$ is a new function which has been first introduced
in \cite{Anikin:2021zxl}.

It is important to notice that the quark spin cannot directly be observed, but we can judge
on the quark spin presence implicitly. As shown in \cite{Anikin:2021zxl}, dispite
the angle $\phi_s$ from
$\epsilon^{- + k_2^\perp s^\perp} = \vec{\bf k}_2^\perp \wedge \vec{\bf s}^{\perp} \sim \sin (\phi_k - \phi_s)$
cannot explicitly be measured in the experiment,
the implementation of the covariant (invariant) integration of $f_{(2)} (y; \,k_2^{\perp\, 2})$
(see Subsec.~\ref{SubSec:Kin-HT-kT} for details) gives the
kinematical constraints on this angle relating the quark spin angle to the corresponding hadron angle.
Namely, the orthogonality condition
required by the covariant integration leads to
the (anti)collinearity of $P^\perp_2$ and $s^\perp$.
As a result, we deal with $\varphi_P = \varphi_s \pm n\pi$
that relates the hadron momentum with the
quark spin vector. In other words, this constraint can be treated as one of conditions for the existence of our new function.

Hence, for the gauge invariant hadron tensor we have the following expression
 \begin{eqnarray}
\label{HadTen-GI-2}
&&{\cal W}_{(1), (2)}^{\mu\nu}= \delta^{(2)}(\vec{\bf q}_\perp)
\int (dx_1)\,(dy) \, \delta(x_1P^{h,+}_1 - q^+)
\nonumber\\
&&\times
\delta(yP^-_2 - q^-)\,
\Big\{
\int (d^2 \vec{\bf k}_2^\perp)
F^{\alpha}_{(1), (2)}(y, \vec{\bf k}_2^\perp)  \,k_{2\, \alpha}^\perp
\Big\}
\nonumber\\
&&
\times
\int (dx_2) \widetilde{B}(x_1,x_2) \,
 \frac{V^\nu_{\perp}}{P^{h}_1\cdot P_2}
\Big[ \frac{P^{h\,\mu}_{1\,\mu}}{y} - \frac{P^{\mu}_{2}}{x_1}\Big] \,,
\end{eqnarray}
where
\begin{eqnarray}
 F^{\alpha}_{(1), (2)}(y, \vec{\bf k}_2^\perp)=
          \begin{pmatrix}
          f_{(1)} (y, \vec{\bf k}_2^\perp) \,\varepsilon^{P^h_1+ \alpha S^\perp} \vspace{0.2cm}\\
          f_{(2)}(y, \vec{\bf k}_2^\perp)  \,  \varepsilon^{P^h_1+ \alpha s^\perp} \vspace{0.2cm}
          \end{pmatrix},
\end{eqnarray}
with the functions $f_{(1)} (y, \vec{\bf k}_2^\perp)$ and $f_{(2)}(y, \vec{\bf k}_2^\perp)$
corresponding to the nucleon polarized and nucleon unpolarized cases.
Then, we focus on the quark axial-vector projection of the nucleon correlator which reads
\begin{eqnarray}
\label{A-me}
\Phi^{[\gamma^- \gamma_5]}(k_2)=\, \varepsilon^{ -+ k^\perp_2 P_2^\perp} \, f_{(3)} (y, \vec{\bf k}_2^\perp)
\end{eqnarray}
and defines the other new $k_\perp$-dependent function \cite{Anikin:2021zxl} .

The covariant (invariant) integration is one of our essential tool (see, for example, \cite{Efimov:1993zg}).
It needs to define the set of exterior Lorentz tensors which are, in our case, $(P_2^\perp, S^\perp, s^\perp)$.
Appendix~\ref{AppC} summarizes the main stages of this technique.

Notice that the covariant integration of $f_{(3)} (y, \vec{\bf k}_2^\perp) k^\perp_{2\,\alpha}$
over $d^2 \vec{\bf k}_2^\perp$ does not contain the vector $P_2^\perp$.
Instead, the tr\-ansverse $k_\perp$-dependence should be realized by the quark spin covariant vector only.
Indeed, the Lorentz covariant integration gives us the following
\begin{eqnarray}
\label{L-c-int}
\int (d^2 \vec{\bf k}_2^\perp)
f_{(3)}(y, \vec{\bf k}_2^\perp)  k_{2\, \alpha}^\perp =  \varepsilon^{\alpha s^\perp + -}
{\mathbb A}\big[ f_{(3)} \big]
\end{eqnarray}
with
\begin{eqnarray}
\label{L-c-int-A}
{\mathbb A}\big[f_{(3)}\big] =
\int (d^2 \vec{\bf k}_2^\perp)
f_{(3)}(y, \vec{\bf k}_2^\perp)
\frac{ \, \varepsilon^{k_{2}^\perp s^\perp + -}}{(s^\perp)^2}.
\end{eqnarray}
Therefore, in the case of axial correlator, the gauge invariant hadron tensor reads
\begin{eqnarray}
 \label{HadTen-GI-2-f3}
&&{\cal W}_{(3)}^{\mu\nu}= \delta^{(2)}(\vec{\bf q}_\perp)
\int (dx_1)\,(dy) \, \delta(x_1P^{h,+}_1 - q^+)\delta(yP^-_2 - q^-)
\nonumber\\
&&\times
\int (dx_2) \widetilde{B}(x_1,x_2)
 |P^\perp_2| \cos\phi_{PS}\,
 \varepsilon^{\nu + P^{h}_1 P^{h}_{1\,\perp}}
 \nonumber\\
&&\times
\frac{1}{P^{h}_1\cdot P_2}\, \Big[ \frac{P^{h\, \mu}_{1}}{y} - \frac{P^{\mu}_{2}}{x_1}\Big]
{\mathbb A}\big[ f_{(3)} \big],
\nonumber\\
\end{eqnarray}
where
$|P^\perp_2| \cos\phi_{PS} = (s^\perp P^\perp_2)$ provided $|s^\perp|=1$.

To conclude this subsection, one remark is in order.
Having restored the corresponding Wilson line in the nucleon correlators,
we may conclude that
the $k_\perp$-dependent parton functions possess the non-trivial properties under
the time-reversal transforms because these transforms
convert the future-pointed WL to the past-pointed WL \cite{Boer:2003cm}.

%%%%%%%%%%%%%%%%%%%%%%%%%%%%
\subsection{Kinematical constraints for extractions of
$k_\perp$-dependent  $f_{(i)}$-functions}
\label{SubSec:Kin-HT-kT}
%%%%%%%%%%%%%%%%%%%%%%%%%%%%

The covariant (invariant) integrations of (\ref{V-me}) and (\ref{A-me}) give the
constraint conditions imposed on the corresponding angles in order to
single out the contribution of given $f_{(i)}$-functions, see Appdendix~\ref{AppC}.

Let us consider the following integration, which appears in the hadron tensor after the factorization
theorem used,
\begin{eqnarray}
\label{V-me-2}
{\cal I}_1=
\int (d^2 \vec{\bf k}_2^\perp)
f_{(1)} (y, \vec{\bf k}_2^\perp) \,
\, \varepsilon^{ -+ S_\perp k^\perp_2 },
\end{eqnarray}
where $\varepsilon^{ -+ S_\perp k^\perp_2 }= \vec{\bf S}_\perp \wedge \vec{\bf k}_2^\perp$.
Hence, we have (we omit the unit vector $\vec{\bf n}$ defining the direction of vector product)
\begin{eqnarray}
\label{V-me-3}
{\cal I}_1=
|\vec{\bf S}_\perp|
\int (d^2 \vec{\bf k}_2^\perp)
f_{(1)} (y, \vec{\bf k}_2^\perp) \,  |\vec{\bf k}_2^\perp| \sin(\phi_S - \phi_{k_2}),
\end{eqnarray}
where, and in what follows, the angle $\phi_A$ is defined in $(\hat x, \hat y)$-plane
as the angle between the vector $A$ and $\hat x$-axis with $A=(S^\perp, s^\perp, P_2^\perp, k_2^\perp)$.
On the other hand, the integration ${\cal I}_1$ can be presented with the help of the covariant integration as
(see Appendix~\ref{AppC})
\begin{eqnarray}
\label{V-me-4}
&&{\cal I}_1=
\int (d^2 \vec{\bf k}_2^\perp)
f_{(1)} (y, \vec{\bf k}_2^\perp)
\\
&&
\times
\Big\{
\, \varepsilon^{ -+ S_\perp P^\perp_2 } \frac{(\vec{\bf k}^\perp_2 \vec{\bf P}^\perp_2)}{(\vec{\bf P}^\perp_2)^2} +
\, \varepsilon^{ -+ S_\perp \alpha}  \varepsilon^{ \alpha s_\perp -+ }
\frac{\varepsilon^{ k^\perp_2 s_\perp -+ }}{\vec{\bf s}_\perp^2}
\Big\}.
\nonumber
\end{eqnarray}
Comparing the representations with and without the covariant integrations, we obtain that
\begin{eqnarray}
\label{V-me-4}
&&
\int (d^2 \vec{\bf k}_2^\perp)
f_{(1)} (y, \vec{\bf k}_2^\perp) \, | \vec{\bf S}_\perp|
\Big\{
\sin(\phi_S - \phi_{k_2}) - \sin(\phi_S - \phi_{P_2})
\nonumber\\
&&
\times \cos(\phi_{P_2} - \phi_{k_2}) -
\sin(\phi_s - \phi_{k_2}) \cos(\phi_{s} - \phi_{S})
\Big\}=0,
\end{eqnarray}
where all angles are determined in the two-dimensional Euclidian space forming the perpendicular plane,
{\it i.e. } $\forall \phi \in \mathbb{R}^2_\perp$.
The equation of (\ref{V-me-4}) has a solution given by
$\phi_{P_2} = \phi_{s} - \pi n$. On the other hand, this condition
can be considered as a necessity condition for the covariant integration.

In the similar manner, we can analyse the other $f_{(i)}$-functions.
Finally, we can fix all the kinematical restrictions on the angles
which allow to extract the corresponding $f_{(i)}$-function contributions to the hadron tensor. We have
\begin{eqnarray}
\label{f-extract}
&&
f_{(1)} (y, \vec{\bf k}_2^\perp) \,\, \text{extracted at} \,\,
\big\{ \phi_{P_2} = \phi_{s} - \pi n \big\};
\\
&&
f_{(2)} (y, \vec{\bf k}_2^\perp) \,\, \text{extracted at} \,\,
\big\{ \phi_{P_2} = \phi_{S} + \pi, \phi_{S} = \phi_{s} - \frac{\pi}{2} \big\};
\nonumber\\
&&
f_{(3)} (y, \vec{\bf k}_2^\perp) \,\, \text{extracted at} \,\,
\big\{ \phi_{S(\text{or} \,\,s)} = \phi_{k_2}, \phi_{s(\text{or} \,\,S)} = \phi_{P_2} \big\}.
\nonumber
\end{eqnarray}
The remaining angles, which are not shown in (\ref{f-extract}), should be considered as free angles.

On the other hand, the kinematical constraints of Eqn.~(\ref{f-extract}) give a possibility to express
the quark spin angle $\phi_{s}$ through the hadron angles $\phi_{P_2}$ and $\phi_{S}$ which
are available in experiments.

%%%%%%%%%%%%%%%%%%%%%%%%%%%%%%%%%%%%%
\subsection{On the complexity of $\widetilde{B}(x_1,x_2)$-function}
\label{SubSec:Comp-B}
%%%%%%%%%%%%%%%%%%%%%%%%%%%%%%%%%%%%%

The complexity of $\widetilde{B}(x_1,x_2)$-function given by
the representation of (\ref{B-fun}) together with (\ref{Conv-Delta}) is very important property
which influences on the observables.
We concentrate on the analysis of complexity and we determine the
possible reasons of its origin.

From Eqns.~(\ref{B-fun}) and (\ref{Conv-Delta}), one can immediately see the first source of complexity which is
due to the gluon pole at $x_2=x_1$ regularized by the complex prescription in the
frame of contour gauge conception \cite{Anikin:2017azc, Anikin:2021osx}.

Then, we discuss the momentum dependence of nonperturbative function related to the correlator.
The exact form of momentum dependence is dictated by the detailization
of interactions in the correlator. Indeed, as an illustrative example,
let us consider the simplest function $\Phi^{[\Gamma]}(k)$ given by
(the symbol of time-ordering is not shown)
\begin{eqnarray}
\label{S-ph-1}
&&\Phi^{[\Gamma]}(k) =
 \int (d^4 z) e^{+i kz} \langle \bar\psi(0) \Gamma \psi(z)  \mathbb{S} \rangle.
\end{eqnarray}
Here, we use the interaction representation with the $\mathbb{S}$-matrix for the correlator.
If we neglect the interactions in the correlator, {\it i.e.} $\mathbb{S}\to \mathbb{I}$, we can see
that all $k$-dependence is accumulated in the exponential:
\begin{eqnarray}
\label{S-ph-2}
&&\Phi^{[\Gamma]}(k) \Big|_{\mathbb{S}=\mathbb{I}} =
 \int (d^4 z) e^{+i kz} \langle \bar\psi(0) \Gamma \psi(z)\rangle.
\end{eqnarray}
Having integrated the function $\Phi^{[\Gamma]}(k)$ over $k^-$ and $\vec{\bf k}_\perp$,
we obtain that the coordinate $z$ of $\psi$-function lies on the light-cone minus direction:
\begin{eqnarray}
\label{S-ph-3}
&& \int (dk^-) (d^2 \vec{\bf k}_\perp)
\Phi^{[\Gamma]}(k^+, k^-, \vec{\bf k}_\perp) \Big|_{\mathbb{S}=\mathbb{I}} =
 \int (d^4 z) e^{+i k^+z^-} \, \delta(z^+) \, \delta^{(2)}(\vec{\bf z}_\perp) \langle \bar\psi(0) \Gamma \psi(z)\rangle.
\end{eqnarray}

However, the momentum dependence becomes more involved if we take into account, say, the second order of
interactions. Namely, replacing $\mathbb{S}$ on $\mathbb{S}^{(2)}$ in Eqn.~(\ref{S-ph-1}), we have
the following expression
\begin{eqnarray}
\label{S-ph-4}
&&\Phi^{[\Gamma]}(k) \Big|_{\mathbb{S}=\mathbb{S}^{(2)}} =
 \int (d^4 z) e^{+i kz} \langle \bar\psi(0) \Gamma \psi(z)
 \Big[ \int (d^4 \xi) \bar\psi(\xi) \hat A(\xi) \psi(\xi)
 \Big]^2
 \rangle =
 \nonumber\\
 &&
  \int (d^4 \xi) e^{+i k \xi} \langle \bar\psi(0) \Gamma
 \Big[ S(k) \int (d^4 \ell) \gamma_\alpha S(k-\ell) \gamma_\beta {\cal D}_{\alpha\beta} (\ell)
 \Big] \psi(\xi)
 \rangle
 \nonumber\\
 &&
 + (\text{the other contributions from Wick's theorem}).
\end{eqnarray}
One can see that in Eqn.~(\ref{S-ph-4}) the $k$-dependence of $\Phi^{[\Gamma]}(k)$ is determined by
the exponential and propagators, cf. Eqn.~(\ref{S-ph-2}).
Thus, neglecting the corresponding interaction, we simplify the momentum dependence of nonperturbative functions.

In this connection, if we neglect the interaction in the channel defined by the light-cone minus direction,
using Eqns.~(\ref{Phi2-func}) and  (\ref{Conv-Delta}) we can get that
\begin{eqnarray}
\label{Phi2-ell-int}
&&\int (d\ell^-) \frac{e^{+i\ell^- \eta_1^+}}{\ell^- - \vec{\ell}^{\,2}_\perp/(2x_{21}P_1^{h\, +}) + i\, \tilde\epsilon}=
\nonumber\\
&&
-2\pi i\,\theta(-\eta_1^+) \,\theta(x_{21})
e^{+i \eta_1^+ \big( \vec{\ell}^{\,2}_\perp/(2x_{21}P_1^{h\, +}) - i\, \tilde\epsilon \big)},
\end{eqnarray}
where $|\vec{\ell}_\perp| \ll P_1^{h\, +}$ has been always assumed.
That is, in this case the Cauchy theorem use provides the complex $i$ accompanying the corresponding residue.
Hence, the $\widetilde{B}$-function takes the form of
\begin{eqnarray}
\label{B-fun-2}
&&\widetilde{B}_{(\text{\tiny appr.})}(x_1,x_2) \equiv \widetilde{\bf B}(x_1,x_2)= \frac{i}{x_2-x_1 - i\epsilon}
\int(d^2\vec{\bf k}_{1}^{\,\perp})  \,
\bar\Phi_{(1)}^{\text{tw-3}}
(x_1; \vec{\bf k}_{1}^{\,\perp})
\nonumber\\
&&
\times
\int(d^2\vec{\ell}_{\,\perp})
\bar\Phi_{(2)}^{\text{tw-2}}(\bar x_2;
\vec{\bf P}_{1}^{h\,\perp} - \vec{\bf k}_{1}^{\,\perp}, \vec{\ell}_{\,\perp}).
\end{eqnarray}

So, we have demonstrated that the origin of complexity of $\widetilde{B}$ is being entirely determined by the
interaction order. Within our approximation, the function $\bar\Phi_{(2)}^{\text{tw-2}}$ contributing to $\widetilde{B}$
has to be treated as a function with the correlator involving the non-interacting quarks.

%%%%%%%%%%%%%%%%%%%%%%%%%
\section{Spin Asymmetry Parameters}
\label{Sec:SSA}
%%%%%%%%%%%%%%%%%%%%%%%%%

In this section, we concentrate on the differential cross sections and
the several spin asymmetry parameters which are available for the experimental studies.

First, we want to mention that, in the most general case, the cross section takes the following standard form
(for the sake of shortness the normalization factors have been absorbed
by the definition of integration measure)
\begin{eqnarray}
\label{gen-x-s-1}
\int d\sigma = \int (d{\mathbb P}{\mathbb S}) \delta^{(4)}\left( \langle mom. conserv.\rangle \right)
\big| {\cal M}\big|^2,
\end{eqnarray}
where the integrated phase space given by $(d{\mathbb P}{\mathbb S})$
expressed through the function $\lambda(s, t, u)$ is the Lorentz invariant object as well as the
process amplitude ${\cal M}$. Therefore, the integrated cross section is a frame-independent function of
Lorentz scalars. Moreover, the contributions of vector components to the given scalars which appear
in the integrated phase space and in the amplitude can be different ones.
Namely, let $s$ be a scalar that parametrizes the phase space and is equal to
$(a\cdot b)_{\text{frame-}1}$ defined by components in the given frame.
The same value of $s$ which is now an argument of the amplitude can be presented by different
product $(a\cdot b)_{\text{frame-}2}$ defined in another frame. It means that the integrated phase space and the
given amplitude as the functions of scalars can be calculated in the different frames depending on our
practical choice. However, the angular dependence of differential cross-sections under the experimental studies
is a frame-dependent subject.

Further, every of spin asymmetries  is given by the hadron tensor that contracted with the lepton tensor.
In our case, we define the lepton tensor as
\begin{eqnarray}
\label{lep-ten}
&&{\cal L}^{\mu\nu} = \big[
\bar u(l_2) \gamma^\nu u(l_1)
\big]^* \big[
\bar u(l_2) \gamma^\mu u(l_1)
\big]=
\nonumber\\
&&
2\left[
l_{1}^{\mu}l_{2}^{\nu} + l_{1}^{\nu}l_{2}^{\mu} - g^{\mu\nu} (l_1 l_2)
\right] +
2 i \lambda_l \epsilon^{\mu\nu l q}
\nonumber\\
&&
\equiv
{\cal L}^{\mu\nu}_U+ i {\cal L}^{\mu\nu}_L,
\end{eqnarray}
where the first term corresponds to the unpolarized leptons while the second term appears
if the initial lepton is longitudinally polarized.

For the present study, the lepton center-of-mass system as known as the CS-frame is
the most convenient frame for both the considered factorization procedure and the analysis of spin asymmetries
related to the gluon pole contributions. Indeed, within the CS-frame,  (a) the initial and final hadron dominant
momenta are separated well ensuring the needed conditions for factorization;
(b) the hadrons possess the nonzero transverse momentum components which are necessary to construct
the observables associated with the gluon poles.
However, from the experimental point of view, the best cho\-ice for the investigation of several observables
with the essential angular dependence is provided by the the initial hadron rest frame.
For example, for the JLab experiments the hadron target at the rest has being used.

In this connection, we can first calculate the contraction of lepton and hadron tensors in terms of Lorentz invariant
scalar products. Afterwards, all scalar products which are appeared in according to the considered case
are expressed through the frame-dependent components which match the preferable phase space system.
We refer the reader to Appendix~\ref{App:Frame} for all details of the frame choice and of the transitions
between the frames.

%%%%%%%%%%%%%%%%%%%%%%%%%%%%
\subsection{Differential cross sections}
\label{SubSec:Diff-XS}
%%%%%%%%%%%%%%%%%%%%%%%%%%%%

Let us now go over to the discussion on the differential cross section.
At the begining, we consider the phase space, we have
the following
\begin{eqnarray}
\label{ps-1}
&&d{\mathbb P}{\mathbb S}=\frac{d^3 \vec{{\bf l}}_2}{(2\pi)^3 2 l_2^0} \frac{d^3 \vec{{\bf P}}^h_1}{(2\pi)^3 2 E^h}=
\nonumber\\
&&
\frac{d^3 \vec{{\bf P}}^h_1}{(2\pi)^3 2 E^h} \int (d^4 q) \delta\left( q^2-2(l_1 q)\right).
\end{eqnarray}
Hence, the differential cross-section reads
\begin{eqnarray}
\label{x-s-2}
&&d\sigma = \frac{d^3 \vec{{\bf P}}^h_1}{(2\pi)^3 2 E^h} \int (d^4 q) \delta\left( q^2-2(l_1 q)\right)
\times
\nonumber\\
&&
{\cal L}_{\mu\nu} T_{\mu\nu}^{\text{GI}}(P^h_1, P_2) {\mathbb F} (x_B, y_B) \delta^{(2)}\left(  \vec{\bf q}_\perp\right),
\end{eqnarray}
where the gauge invariant hadron tensor, see (\ref{HadTen-GI}), has been presented in the form of
\begin{eqnarray}
\label{HadTen-GI-2-2}
{\cal W}_{\mu\nu} = T_{\mu\nu}^{\text{GI}}(P^h_1, P_2) {\mathbb F} (x_B, y_B) \delta^{(2)}\left(  \vec{\bf q}_\perp\right)
\end{eqnarray}
with the gauge invariant algebraic tensor given by
\begin{eqnarray}
\label{GI-ten}
T_{\mu\nu}^{\text{GI}}(P^h_1, P_2)=
\frac{V^\perp_\nu}{P^{h}_1\cdot P_2}
\Big[ \frac{P^{h}_{1\,\mu}}{y} - \frac{P_{2\, \mu}}{x_1}\Big]
\end{eqnarray}
and
\begin{eqnarray}
\label{Had-Ten-F}
&&{\mathbb F} (x_B, y_B) =
\int (dx_1)\,(dy) \, \delta(x_1P^{h,+}_1 - q^+)\delta(yP^-_2 - q^-)
\nonumber\\
&&\times
\Big\{ \int (d^2 \vec{\bf k}_{2}^\perp) F(y; \vec{\bf k}_{2}^\perp) w\big(\vec{\bf k}_{2}^{\perp\,2} \big)
\Big\}
\int (dx_2) \widetilde{B}(x_1,x_2).
\end{eqnarray}
The functions $F(y; \vec{\bf k}_{2}^\perp)$ and $w\big(\vec{\bf k}_{2}^{\perp\,2})$ denote the
corresponding distribution function and the weight function

We now assume that
the contraction of (\ref{lep-ten}) and (\ref{HadTen-GI-2-2}) is being expressed in terms of Lorentz scalars.
The two-dimension delta function $\delta^{(2)}\left(  \vec{\bf q}_\perp\right)$ of Eqns.~(\ref{HadTen-GI-2}) and/or
(\ref{HadTen-GI-2-2}) is appeared thanks for the decomposition of (\ref{delta-decom}).

Introducing the photon rapidity as
\begin{eqnarray}
\label{ph-rap}
2\, y_\gamma = \ln \frac{q^+}{q^-},
\end{eqnarray}
the differential cross section takes the form of
\begin{eqnarray}
\label{x-s-3}
d\sigma = \frac{d^3 \vec{{\bf P}}^h_1}{(2\pi)^3 2 E^h} \int d y_\gamma
{\cal L}_{\mu\nu} T_{\mu\nu}^{\text{GI}}(P^h_1, P_2) {\mathbb F} (x_B, y_B) \Big |_{\vec{\bf q}_\perp=0}
\end{eqnarray}
or
\begin{eqnarray}
\label{x-s-4}
(2\pi)^3 d\sigma = \frac{d\tilde z}{\tilde z}
d^2 \vec{{\bf P}}^{h\, \perp}_1
\int d y_\gamma
{\cal L}_{\mu\nu} T_{\mu\nu}^{\text{GI}}(P^h_1, P_2) {\mathbb F} (x_B, y_B) \Big |_{\vec{\bf q}_\perp=0}
\end{eqnarray}
where
\begin{eqnarray}
\label{tilde-z}
&&\frac{d (P^h_1)_3}{E^h_1} \approx \frac{d (P^h_1)_3}{(P^h_1)_3} \approx \frac{d P^{h\,+}_1}{P^{h\,+}_1} = \frac{d\tilde z}{\tilde z},
\nonumber\\
&&
\tilde z = \frac{P_2\cdot P^{h}_1}{P_2\cdot q} \Big | _{\text{CS}}=\frac{P^{h\,+}_1}{q^+}.
\end{eqnarray}

Following \cite{Barone:2001sp}, we can also introduce the other fracture parameter as
\begin{eqnarray}
\label{tilde-y}
\tilde y = \frac{P_2\cdot q}{P_2\cdot l_1} \Big | _{\text{CS}}=\frac{q^+}{l^+_1},
\end{eqnarray}
and we can readily derive that
\begin{eqnarray}
\label{tilde-y-2}
1-\tilde y = e^{2(y_\gamma-y_\ell)}
\end{eqnarray}
with the lepton rapidity $y_\ell$. Hence, if $y_\gamma\to 0$ (see Eqn.~(\ref{ph-rap})) then
\begin{eqnarray}
\label{tilde-y-3}
1-\tilde y = e^{-2 y_\ell} \Rightarrow d\tilde y = 2 e^{-2 y_\ell} dy_\ell.
\end{eqnarray}
For the region of small lepton rapidity, we get that $d\tilde y \approx 2dy_\ell$.
The photon momentum integration in the phase space is very crucial
to complete the factorization procedure for the process under our consideration.

%%%%%%%%%%%%%%%%%%%%%%%%%%%%%%%%%%
\subsection{The case of unpolarized leptons and hadrons}
\label{SubSec:Unp}
%%%%%%%%%%%%%%%%%%%%%%%%%%%%%%%%%%

Since every of single spin asymmetries has to be normalized by the unpolarized cross section,
we begin with a consideration of the case where both the lepton and hadron tensors correspond
to the unpolarized particles. In Section~\ref{Sec:Comp-LO-NLO},
it has been proven that if we focus on the particular kinematics where $x\to 1$ and the intermediate (anti)quark is massless one,
the leading order hadron tensor is suppressed in comparison with the next-to-leading hadron tensor related to the
gluon pole contributions.

Hence, the differential cross section is given by the
following contractions of lepton and hadron tensors:
\begin{eqnarray}
\label{xsec-unpl}
 d\sigma^{unpol.} \sim {\cal L}^{U}_{\mu\nu}
 \Big\{
 {\cal W}^{(2)}_{\mu\nu} \big( \Re\text{e} \widetilde{\bf B}\big) +
  {\cal W}^{(3)}_{\mu\nu} \big(  \Im\text{m}  \widetilde{\bf B}\big)
 \Big\},
\end{eqnarray}
where the hadron tensor is presented by (\ref{HadTen-GI-2}) and (\ref{HadTen-GI-2-f3}), while
the lepton tensor is taken from (\ref{lep-ten}).

In Eqn.~(\ref{xsec-unpl}), the tensor contractions give the following expressions
 \begin{eqnarray}
\label{xsec-unp-2-2}
&&
{\cal L}^{U}_{\mu\nu}
 {\cal W}^{(2)}_{\mu\nu} \big(  \Re\text{e}\widetilde{\bf B}\big)=
\nonumber\\
&&
\int (dx_1)\,(dy) \, \delta(x_1P^{h,+}_1 - q^+)
\delta(yP^-_2 - q^-)
\nonumber\\
&&\times
\Big\{
\int (d^2 \vec{\bf k}_2^\perp)
 f_{(2)}(y, \vec{\bf k}_2^\perp)
\varepsilon^{P^{h}_1 + k_2^\perp s^\perp}
\Big\}
\nonumber\\
&&
\times
\int (dx_2)\Re\text{e} \widetilde{\bf B}(x_1,x_2) \,
 \frac{(\hat x \cdot P^{h\, \perp}_1)}{P^{h}_1\cdot P_2}
 \sin 2\theta \, \cos\varphi ,
\end{eqnarray}
and
 \begin{eqnarray}
\label{xsec-unp-2-3}
&&
{\cal L}^{U}_{\mu\nu}
  {\cal W}^{(3)}_{\mu\nu} \big( \Im\text{m} \widetilde{\bf B}\big) =
\nonumber\\
&&
\int (dx_1)\,(dy) \, \delta(x_1P^{h,+}_1 - q^+)
\delta(yP^-_2 - q^-)
\nonumber\\
&&\times
\Big\{
\int (d^2 \vec{\bf k}_2^\perp)
 f_{(3)}(y, \vec{\bf k}_2^\perp)
\varepsilon^{ P^{h}_1 + k_2^\perp P_2^\perp}
\Big\}
\nonumber\\
&&
\times
\int (dx_2)\Im\text{m}\widetilde{\bf B}(x_1,x_2) \,
 \frac{\hat y \wedge P^{h}_{1\, \perp}}{m_N\, P^{h}_1\cdot P_2}
 \sin 2\theta \, \sin\varphi.
\end{eqnarray}
We remind once more that the angular dependences have been fixed in the CS-frame, see
Appendix~\ref{App:Frame}.

%%%%%%%%%%%%%%%%%%%%%%%%%%%
\subsection{Single spin asymmetry}
\label{SubSec:SSA-2}
%%%%%%%%%%%%%%%%%%%%%%%%%%%

The single spin asymmetry can be determined for our process with unpolarized leptons provided
the (target) initial hadron has a transverse polarization.
The practical reason for the experimental study of this asymmetry can be formulated as the following:
the given asymmetry directly probes the gluon pole, {\it i.e.}
if the asymmetry is nonzero, the gluon pole exists.
Indeed, the single spin asymmetry is usually
related to the imaginary part of the gluon pole which is encoded in the
distribution function $\widetilde{B}(x_1,x_2)$ (see, the prefactor $1/(x_2-x_1 - i \epsilon)$ of Eqn.~(\ref{B-fun})).
Notice that, within our approximation described in subsection~\ref{SubSec:Comp-B}, the single spin asymmetry is given by
$\Re\text{e} \widetilde{\bf B}\sim \delta(x_2-x_1)$.

Thus, we have
\begin{eqnarray}
\label{SSA-1}
 {\cal A}_{UT} \sim {\cal L}^{U}_{\mu\nu}
 \Big\{
  {\cal W}_{\mu\nu} \big( \Re\text{e} \widetilde{\bf B}\big) +
 {\cal W}^{(1)}_{\mu\nu} \big( \Re\text{e} \widetilde{\bf B}\big),
 \Big\}
\end{eqnarray}
where the hadron tensor is determined by Eqns.~(\ref{HadTen-GI}) and (\ref{HadTen-GI-2}).
The expression for this contraction reads
 \begin{eqnarray}
\label{xsec-unp-3-1}
&&
{\cal L}^{U}_{\mu\nu}
 \Big\{
  {\cal W}_{\mu\nu} \big( \Re\text{e} \widetilde{\bf B}\big) +
 {\cal W}^{(1)}_{\mu\nu} \big( \Re\text{e} \widetilde{\bf B}\big),
 \Big\}
=
\nonumber\\
&&
\int (dx_1)\,(dy) \, \delta(x_1P^{h,+}_1 - q^+)
\delta(yP^-_2 - q^-) \int (dx_2)
\nonumber\\
&&
\times
\Re\text{e}\widetilde{\bf B}(x_1,x_2)
\Big[
 \frac{(\hat x \cdot P^{h\, \perp}_1)}{P^{h}_1\cdot P_2}
 \sin 2\theta \, \cos\varphi
 \nonumber\\
 &&
 \times
 \Big\{
 f_T(y)
 \varepsilon^{P^h_1+S^\perp P_2^\perp} +
\int (d^2 \vec{\bf k}_2^\perp)
 f_{(1)}(y, \vec{\bf k}_2^\perp)
\varepsilon^{P^{h}_1 + k_2^\perp S^\perp}
\Big\}
\nonumber\\
&&
+
m_N\, h_1(y)
\hat y \wedge S_{\perp}
 \sin 2\theta \, \sin\varphi
\Big].
\end{eqnarray}

%%%%%%%%%%%%%%%%%%%%%%%%%%
\subsection{Double spin asymmetry}
\label{SubSec:D-SSA}
%%%%%%%%%%%%%%%%%%%%%%%%%%

It is important to show that
the double spin asymmetry, associated with the longitudinally polarized initial lepton
and the transverse polarized initial hadron, does not exist within our approach.
Indeed, the double spin asymmetry is related to the following contraction between
the lepton and hadron tensors
\begin{eqnarray}
&&{\cal A}_{LT} \sim {\cal L}^{L}_{\mu\nu}
 \Big\{
  {\cal W}_{\mu\nu} \big(  \Im\text{m} \widetilde{\bf B}\big) +
 {\cal W}^{(1)}_{\mu\nu} \big( \Im\text{m} \widetilde{\bf B} \big),
 \Big\}
\nonumber\\
&&
\sim 2 \lambda_l
\epsilon^{\mu \nu l q}  \Big[ \frac{P^{h}_{1\, \mu}}{y} - \frac{P_{2\, \mu}}{x_1}\Big]
\Big\{
P^{h\, \perp}_{1\, \nu}  \varepsilon^{P^h_1+S_\perp P_2^\perp} \oplus\,
m_N \varepsilon^{\nu S_\perp P_2 P^h_1}_\perp
\Big\}
\nonumber\\
&&
\sim \epsilon^{\hat z \nu l q}
\big\{ P^{h\, \nu}_{1\, \perp} \oplus \varepsilon^{\nu S_\perp P_2 P^h_1}_\perp
\big\}
= (\hat z \cdot q)
\vec{\bf l}_\perp \wedge \vec{\bf T}_{\perp} = 0
\end{eqnarray}
due to the gauge invariant combination $(\hat z \cdot q) = 0$.

%%%%%%%%%%%%%%%%%%%
\section{Conclusions}
\label{Sec:Concl}
%%%%%%%%%%%%%%%%%%%

In the paper, we have used the approach
which is ground on the archetypal factorization procedure established in
a series of seminal papers
\cite{Efremov:1976ih,  Efremov:1978rn, Efremov:1978xm, Efremov:1978cu}.
This approach has been applied to the semi-exclusive mode of the hadron production processes
where the hadron tensor is given by the
corresponding convolution of the distribution function and the
distribution amplitudes provided a large value of intermediate quark fraction, $x\to 1$.
We have singled out the fact that, in our approach,
the transverse momenta of photon and produced hadron
are not tied by any conditions
in contrast to the TMD-factorization approaches.
This is definitely a preponderance of our approach which gives a possibility to
obtain a new single spin asymmetry associated with the gluon pole contribution
in a wide region of the transverse momentum of produced hadron.
We have suggested to consider this kind of asymmetry as a object of experimental measurements
at a large value of $\vec{\bf P}^{h}_\perp$ using the existing data of JLab.

In the paper, we have demonstrated that
the leading order hadron tensor is suppressed by the model-independent kinematical factor
$(\bar x_1 P_1^{h, +})^2/m_q$, see Eqn.~(\ref{E-lc}), for $x_1\to 1$ and $m_q\to 0$
in comparison with the
gluon pole contribution to the hadron tensor associated with the gluon radiations.
Based on this finding, we have investigated the manifestation of gluon poles in the lepton-hadron
collision of hadron production
at large $x$, where the produced hadron can be described by the corresponding distribution amplitude.

In the frame of the contour gauge conception, it is known that the twist three distribution function \cite{Anikin:2021osx}
which arises from the quark-gluon correlators have to be treated as a complex functions due to the gluon poles.
In this context, we have studied the different sources of functional complexity
which are associated with the gluon pole effects.

We have also introduced the new transverse momentum dependent
distribution functions, see Eqns.~(\ref{V-me}) and (\ref{A-me}), which should extend our knowledge on the hadron spin structure
\cite{Anikin:2021zxl}.
We have demonstrated that these new functions together with
the gluon pole contributions have formed the new observables which should be accessible in
the experimental studies based on the recent data of JLab.

%%%%%%%%%%%%%%%%%%%%%%%%%%%%%%%%%%%%%%%%%%%%%%%%%%%%%%%%%%%%%%%%%%%%%%%%%%%%%%%%%%

\acknowledgments

We thank colleagues from the Theoretical Physics Division of NCBJ (Warsaw)
for useful and stimulating discussions. Our special thanks go to Prof. Harut Avakian of JLab.
This work is supported by the Ulam Program of NAWA No.
PPN/ULM/2020/1/00019.
The work of L.Sz. is supported by the grant 2019/33/B/ST2/02588  of the National Science Centre in Poland.

%%%%%%%%%%%%%%%%%%%%%%%%%%%%%%%%%%%%%%%%%%%%%%%%%%%%%%%%%%%%%%%%%%%%%%%%%%%%%
%%%%%%%%%%%%%%%%%%%%%%%%%%%%%   Appendix   %%%%%%%%%%%%%%%%%%%%%%%%%%%%%%%%%%
%%%%%%%%%%%%%%%%%%%%%%%%%%%%%%%%%%%%%%%%%%%%%%%%%%%%%%%%%%%%%%%%%%%%%%%%%%%%%
\appendix
\renewcommand{\theequation}{\Alph{section}.\arabic{equation}}
%\section*{Appendix}
%%%%%%%%%%%%%%%%%%%%%%%%%%%%%%%%%%%%%%%%%%%%%%%%%%%%%%%%%%%%%%%%%%%%%%%%%%%%%%%%%%%%%%%%%%

%%%%%%%%%%%%%%%%%%%%%%%%
\section{On the frame choice}
\label{App:Frame}
%%%%%%%%%%%%%%%%%%%%%%%%

In this Appendix, we present the Lorentz transforms which are tying the CS-frame and the initial hadron rest frame.
The boost for the initial hadron can be performed by two steps: (a) the Lorentz rotation in
$\left( (P_{2})_ 0, (P_{2})_1 \right)$-plane
and, then, (b) the Lorentz rotation in $\left( (P^{\prime}_{2})_0, (P_{2})_3 \right)$-plane.
As a result, we have
\begin{eqnarray}
\label{L-boost}
P_{2\,\mu} \Big|_{\text{RS}}
= {\mathbb B}_{\mu\nu}\, P_{2\,\nu} \Big|_{\text{CS}}
\end{eqnarray}
where
\begin{eqnarray}
\label{B-matrix}
&&{\mathbb B}_{00}=\cosh\theta_2 \cosh\theta_1, {\mathbb B}_{01}=\cosh\theta_2 \sinh\theta_1,
\nonumber\\
&&
{\mathbb B}_{02}=0, {\mathbb B}_{03}=\sinh\theta_2;
\nonumber\\
&&
{\mathbb B}_{10}=\sinh\theta_1, {\mathbb B}_{11}=\cosh\theta_1,
{\mathbb B}_{12}=0, {\mathbb B}_{13}=0;
\nonumber\\
&&{\mathbb B}_{20}=0, {\mathbb B}_{21}=0,
{\mathbb B}_{22}=1, {\mathbb B}_{23}=0;
\nonumber\\
&&{\mathbb B}_{30}=\sinh\theta_2\cosh\theta_1, {\mathbb B}_{31}=\sinh\theta_2\sinh\theta_1,
\nonumber\\
&&
{\mathbb B}_{32}=0, {\mathbb B}_{33}=\cosh\theta_2;
\end{eqnarray}
with
\begin{eqnarray}
\label{L-angles}
2\theta_1=\ln \frac{(P_{2})_0 - (P_{2})_1}{(P_{2})_0 + (P_{2})_1},\quad
2\theta_2=\ln\frac{(P^{\prime}_{2})_0 - (P_{2})_3}{(P^{\prime}_{2})_0 + (P_{2})_3}
\end{eqnarray}
and
\begin{eqnarray}
\label{E-boost}
(P^{\prime}_{2})_0=(P_{2})_0\cosh\theta_1 + (P_{2})_1 \sinh\theta_1.
\end{eqnarray}
From Eqn.~(\ref{L-angles}), we can readily obtain that
\begin{eqnarray}
\label{B-cond}
(P_{2})_0 \not= \pm (P_{2})_1, \quad (P^{\prime}_{2})_0 \not= (P_{2})_3,
\quad P_{2}^2 = M_2^2 \not= 0.
\end{eqnarray}

Having used the hadron momentum representations in CS-frame given by
(cf. (\ref{Sudakov-decom-1}))
\begin{eqnarray}
\label{Had-Mom-CS}
&&P_1^h=\left(
E_1^h, |\vec{\bf P}^h_1|\sin\alpha, 0, |\vec{\bf P}^h_1|\cos\alpha
\right)
\nonumber\\
&&
P_2=\left(
E_2, |\vec{\bf P}_2|\sin\alpha, 0, - |\vec{\bf P}_2|\cos\alpha
\right),
\end{eqnarray}
we derive that
\begin{eqnarray}
\label{boost-ang}
&&2 \theta_1=\ln \frac{1- \beta_2 \sin\alpha}{1+ \beta_2 \sin\alpha}, \,
\nonumber\\
&&
2 \theta_2= \ln \frac{\sqrt{1-\beta_2^2 \sin^2\alpha} + \beta_2\cos\alpha}{\sqrt{1-\beta_2^2 \sin^2\alpha} - \beta_2\cos\alpha},
\end{eqnarray}
where
\begin{eqnarray}
\label{R-par}
\beta_2=|\vec{\bf P}_2|/E_2, \quad P_2^{\perp\, 2} \approx M^2_2 \ll \tilde Q^2.
\end{eqnarray}
Hence, we have
\begin{eqnarray}
\label{boost-ang-2}
&&\cosh \theta_1=\frac{1}{\sqrt{1-\beta_2^2 \sin^2\alpha}},
\sinh \theta_1= -  \frac{\beta_2 \sin\alpha}{\sqrt{1-\beta_2^2 \sin^2\alpha}},
\nonumber\\
&&
\cosh \theta_2= \sqrt{\frac{1 - \beta_2^2 \sin^2\alpha}{1-\beta_2^2}},
\sinh \theta_2=\frac{ \beta_2 \cos\alpha }{ \sqrt{1-\beta_2^2} }.
\end{eqnarray}

So, we adhere the system where the factorization of hadron tensor is more simple from the theoretical point
of view. As mentioned, this system can be traced from the CS-frame (the lepton center-of-mass system).
Then, we make a transformation to the initial hadron rest
system which is more suitable for the considered experiment.

Within the CS-frame, the lepton sector is determined as
\begin{eqnarray}
\label{lepmom}
&&2 l_{1\,\mu} = Q\, L_\mu(\theta,\varphi; \hat x, \hat y, \hat z) + q_\mu,
\nonumber\\
&&
2 l_{2\,\mu} = Q\,L_\mu(\theta,\varphi; \hat x, \hat y, \hat z) - q_\mu
\end{eqnarray}
with $-q^2=2 (l_1 l_2)= Q^2$ and
\begin{eqnarray}
&&L_\mu(\theta,\varphi; \hat x, \hat y, \hat z)=
\nonumber\\
&&
\hat x_\mu\, \cos\varphi \,\sin\theta +
\hat y_\mu\, \sin\varphi \,\sin\theta + \hat z_\mu\, \cos\theta,
\end{eqnarray}
where the normalized axes are determined as $\hat x_\mu= x_\mu/\sqrt{-x^2}$ and so on.
In this system, we have the basis formed by  (here, symbol $\sim$ means ``behaves as" and
the hat symbol means the corresponding normalization)
\begin{eqnarray}
\label{basis-lep}
&&\hat x_\mu\sim \hat q^\perp_\mu\sim \hat e^x_\mu=(0,1, 0, 0),
\nonumber\\
&&
\hat y_\mu\sim \epsilon^\perp_{\mu \hat q}\sim \hat e^y_\mu=(0, 0, 1, 0),
\nonumber\\
&&\hat z_\mu\sim \hat P^h_{1\,\mu} - \hat P_{2\, \mu}.
\end{eqnarray}
After the boost transforms from CS-frame to the RS-frame defined by (\ref{B-matrix}),
the scalar product $l_1\cdot \hat z$ takes the following form
\begin{eqnarray}
\label{lz-RS}
l_1\cdot \hat z \Big|_{RS} = {\mathbb A}(\alpha) +
{\mathbb B}(\alpha) \cos\varphi \sin\theta + {\mathbb C}(\alpha) \cos\theta,
\end{eqnarray}
where
\begin{eqnarray}
\label{A-B-C-ff}
&&{\mathbb A}(\alpha) = \frac{Q \beta_2\cos\alpha}{1-\beta_2^2},\quad
{\mathbb B}(\alpha) = - \frac{Q \beta_2^2 \sin2\alpha}{2(1-\beta_2^2)},
\nonumber\\
&&
{\mathbb C}(\alpha) = \frac{Q (1+ \beta_2^2 \cos2\alpha)}{2(1-\beta_2^2)}.
\end{eqnarray}

%%%%%%%%%%%%%%%%%%%%%%%%%%%%%%%%%%
\section{The alternative representation of functions with the
on-shell fields in correlators}
\label{AppA}
%%%%%%%%%%%%%%%%%%%%%%%%%%%%%%%%%%

Now, we give an alternative representations for the functions which contain the
correlators with the on-shell fields. As shown in section~\ref{SubSec:LO-Function},
the leading order function $\bar\Phi(k_1)$, which stems from the squared amplitude,
is presented in the product form of functions $\bar\Phi^{(1)}_{\text{o}}(k_1)$ and $\bar\Phi^{(2)}_{\text{o}}(k_1)$.
Besides this representation, it is convenient to present the function $\bar\Phi(k_1)$ in the equivalent form as
\begin{eqnarray}
\label{barPhi-Ap-1}
\bar\Phi(k_1) = \frac{E^2}{m_q}\, \bar\Phi^{(2)}(P^h_1-k_1)\,
\bar\Phi^{(1)}(k_1)\delta\Big(
\frac{(k_1-P_1^{h})^2}{m_q^2} - 1
\Big)
\end{eqnarray}
where
\begin{eqnarray}
\label{Phi-Func-Ap-2}
\bar\Phi^{(1)}(k_1)=
\int (d^4 \xi_1) e^{ik_1\xi_1}
\langle 0 |  \bar\psi(\xi_1)  \psi(0) | P_1^h \rangle
\end{eqnarray}
and
\begin{eqnarray}
\label{Phi-Func-Ap-3}
&& \bar\Phi^{(2)}(P^h_1-k_1)=
\nonumber\\
&&
\int (d^4 \eta_1) e^{-i(P^h_1-k_1)\eta_1}
\langle P_1^h |  \bar\psi(\eta_1)  \psi(0) | 0 \rangle.
\end{eqnarray}
 That is, the functions $\bar\Phi^{(1)}(k_1)$ and $ \bar\Phi^{(2)}(P^h_1-k_1)$ involve now the correlators with
 only the off-shell fields and, at the same time, the on-shellness of the intermediate (anti)quark with the momentum
 $k_1-P_1^{h}$ is ensured by the delta function.

For the use of the limits given by $m_q\to 0$ and $x_1\to 1$, the function  $\bar\Phi(k_1)$ can be written as
\begin{eqnarray}
\label{barPhi-Ap-2}
&&\bar\Phi(k_1) =
\frac{m_q}{4 \bar x_1 P_1^{h\, +}}\,
\big[
 \bar x_1 P_1^{h\, +} + k_1^-
\big]^2
\\
&&
\times
\delta\Big(
k_1^- - \frac{(\vec{\bf k}^\perp_1-\vec{\bf P}_1^{h\, \perp})^2+m_q^2}{2\bar x_1 P_1^{h\, +}}
\Big)
\bar\Phi^{(2)}(P^h_1-k_1)\,
\bar\Phi^{(1)}(k_1).
\nonumber
\end{eqnarray}
It is important to emphasize that the leading order function $\bar\Phi(k_1)$ exists only if
the intermediate (anti)quark is on-shell. In other words, the function $\bar\Phi(k_1)$
can be extended beyond the on-shell condition and it is uniquely
related to the square or interference of two amplitudes which are given by function
$\bar\Phi^{(1)}_{\text{o}}(k_1)$ and $ \bar\Phi^{(2)}_{\text{o}}(P^h_1-k_1)$.
In this connection, the on-shell delta function in Eqns.~(\ref{barPhi-Ap-1}) and (\ref{barPhi-Ap-2}) is mandatory
for the representations.

Let us now go over to the alternative representations for the next-to-leading function $\bar\Phi^{(A)}(k_1, \ell)$.
We remind that in section~\ref{SubSec:NLO-Function} the representation for this function is obtained by the
direct extension of the leading order function $\bar\Phi(k_1)$.
Analogously to the above-mentioned way, there is also the equivalent and alternative
representation of  $\bar\Phi^{(A)}(k_1, \ell)$
which is more convenient to use in the most cases. In order to demonstrate it, we  begin with
(the spin $\lambda$-indices are omitted)
\begin{eqnarray}
\label{NLO-fun-1}
&&\bar\Phi(k_1) = \int (d^4 \xi)  e^{-ik_1\xi}\, \int\frac{d^3 \vec{{\bf K}}}{(2\pi)^3 2 K_0}
\nonumber\\
&&
\times
\langle 0|  \bar\psi(0)  d^+(K) | P_1^h\rangle \, \langle P_1^h | d^-(K) \mathbb{S} \psi(\xi) | 0 \rangle^I
\end{eqnarray}
written in the interaction representation.
The next stage is a commutation of $\mathbb{S}$-matrix with the operator $d^-(K)$ given the following expression in the
first order of interaction
\begin{eqnarray}
\label{NLO-fun-2}
&&\hspace{-0.7cm}\bar\Phi^{(A)}(k_1) =
\int (d^4 \xi)  e^{-ik_1\xi}\,
\int\frac{d^3 \vec{{\bf K}}}{(2\pi)^3 2 K_0}
\int (d^4\eta) e^{+iK\eta}
\nonumber\\
&&\hspace{-0.7cm}
\times
\langle 0|  \bar\psi(0)  d^+(K) | P_1^h\rangle \,
\langle P_1^h | \big[
\bar\psi(\eta) \hat A(\eta) v(K)\big]
 \psi(\xi) | 0 \rangle^I.
\end{eqnarray}
Having used that
\begin{eqnarray}
\label{Int-K}
\int\frac{d^3 \vec{{\bf K}}}{(2\pi)^3 2 K_0}
= \int (d^4 K) \delta\big( K^2 - m_q^2\big)
\end{eqnarray}
together with Eqn.~(\ref{q-on}), after some algebra we derive
\begin{eqnarray}
\label{NLO-fun-3}
\bar\Phi^{(A)}(k_1) =
\bar\Phi^{(2)}(k_1) \hat A(0) \bar\Phi^{(1)}_{\text{o}}(k_1-P^{h}_1)
\end{eqnarray}
where
\begin{eqnarray}
\label{Phi2-func-A-4}
\bar\Phi^{(2)}(k_1; \ell)=
 \int (d^4 z)\, e^{- ik_1z}\, \langle P_1^h | \bar\psi(0) \,\psi(z) | 0 \rangle\,,
\end{eqnarray}
\begin{eqnarray}
\label{Phi1-func-A-5}
\hspace{-0.3cm}\bar\Phi^{(1)}_{\text{o}}(k_1)= \int (d^4\xi)e^{-i (k_1- P_1^h)\xi}
\langle 0 | \bar\psi(0) \psi_{\text{on-sh.}}(\xi) | P_1^h \rangle.
\end{eqnarray}
In Eqn.~(\ref{NLO-fun-3}) $A_\alpha(0)$ denotes the gluon field operator in the $x$-space (the position space).
This gluon operator gives the gluon propagator in the hadron tensor after paring with the other
gluon field operator appearing in the radiation correction to the corresponding quark field, see Eqns.~(\ref{HadTen-St}) and
(\ref{HadTen-NonSt}).

In analogy with the leading order function, see Eqn.~(\ref{barPhi-Ap-1}), the function $\bar\Phi^{(A)}(k_1)$ can be presented in
the form as
\begin{eqnarray}
\label{NLO-fun-6}
&&\bar\Phi^{(A)}(k_1) =
\bar\Phi^{(2)}(k_1) \hat A(0) \bar\Phi^{(1)}(k_1-P^{h}_1)
\nonumber\\
&&\times
\delta\Big(
\frac{(k_1-P_1^{h})^2}{m_q^2} - 1
\Big)
\end{eqnarray}
or
\begin{eqnarray}
\label{NLO-fun-7}
&&\bar\Phi^{(A)}(k_1) =
\bar\Phi^{(2)}(k_1) \hat A(0) \bar\Phi^{(1)}(k_1-P^{h}_1)
\nonumber\\
&&\times
\frac{m_q^2}{2 \bar x_1 P_1^{h\, +}}
\delta\Big(
k_1^- - \frac{(\vec{\bf k}^\perp_1-\vec{\bf P}_1^{h\, \perp})^2+m_q^2}{2\bar x_1 P_1^{h\, +}}
\Big).
\end{eqnarray}

Notice that in contrast to the leading order function, the next-to-leading function $\bar\Phi^{(A)}(k_1)$
allow the off-shell extension due to the presence the interaction vertex between  $\bar\Phi^{(2)}(k_1)$
and $\bar\Phi^{(1)}(k_1-P^{h}_1)$.

%%%%%%%%%%%%%%%%%%%%%%%%%%%%
\section{On the off-shell extension}
\label{AppB}
%%%%%%%%%%%%%%%%%%%%%%%%%%%%

In the Appendix, we give a short description of the off-shell extension method.
The necessity of the off-shell extension is dictated by the fact that, as a rule,
the corresponding parton distributions stem from the factorization procedure
applied to the amplitudes involving the non-perturbative correlators with
the off-shell fields.

The principal basis of our method is a statement that the permutation relations for fields are
the translation invariant subjects for both the on-shell and off-shell particles.
We remind that this general statement is forming one of the fundament axioms.

For a pedagogical reason, let us start from the permutation relations written for the on-shell scalar particles. We have
\begin{eqnarray}
\label{B-1}
[\phi^+_{\text{on-sh.}}(x), \, \phi^-_{\text{on-sh.}}(y)] = D^+(x, y).
\end{eqnarray}
In QFT, it is imposed that the function $D^+$ must depend on the difference $x-y$ only, {\it i.e.}
\begin{eqnarray}
\label{B-2}
D^+(x, y) = D^+(x-y).
\end{eqnarray}
Since the Fourier transforms are the linear operations,
the translation invariance property of the permutation relations, see Eqn.~(\ref{B-2}), is determining the commutation relations
for the Fourier images of $\phi^{\pm}$. Indeed,  having used the Fourier transforms, we have the following
\begin{eqnarray}
\label{B-3}
&&[\phi^+_{\text{on-sh.}}(x), \, \phi^-_{\text{on-sh.}}(y)] =
\nonumber\\
&&
\int (d^3 \vec{\bf k}) (d^3 \vec{\bf p})
e^{ikx - ipy}  [\tilde\phi^+_{\text{on-sh.}}(\vec{\bf k}), \, \tilde\phi^-_{\text{on-sh.}}(\vec{\bf p})],
\end{eqnarray}
where
\begin{eqnarray}
\label{Ft-on-sh-1}
&&\tilde\phi^\pm_{\text{on-sh.}}(\vec{\bf k}) \equiv \frac{a^\pm_{\text{on-sh.}} (k)}{2E(\vec{\bf k})} \equiv
\frac{a^\pm_{\text{on-sh.}} (\vec{\bf k})}{\sqrt{2E(\vec{\bf k})}},
\nonumber\\
&&E(\vec{\bf k})=\sqrt{\vec{\bf k}^2 + m^2}.
\end{eqnarray}
From Eqn.~(\ref{B-3}), one can see that in order to ensure the translation invariance of the permutation relation,
we have to demand the commutation relation $[\tilde\phi^+_{\text{on-sh.}}(\vec{\bf k}), \, \tilde\phi^-_{\text{on-sh.}}(\vec{\bf p})]$
to be proportional to the three-dimensional delta function, {\it i.e.}
\begin{eqnarray}
\label{B-4}
 [\tilde\phi^+_{\text{on-sh.}}(\vec{\bf k}), \, \tilde\phi^-_{\text{on-sh.}}(\vec{\bf p})] = \frac{1}{2E(\vec{\bf k})}
 \delta^{(3)}(\vec{\bf k} - \vec{\bf p}).
\end{eqnarray}
Hence, we obtain the standard Paul-Jordan function given by
\begin{eqnarray}
\label{B-5}
\hspace{-0.3cm}[\phi^+_{\text{on-sh.}}(x), \, \phi^-_{\text{on-sh.}}(y)] =
\int \frac{(d^3 \vec{\bf k})}{2E(\vec{\bf k})}
e^{ik(x - y)} \equiv D^+(x-y).
\end{eqnarray}

Notice that the Fourier transforms of on-shell fields, see Eqn.~(\ref{B-3}), can be presented in the four-dimensional forms as
\begin{eqnarray}
\label{B-6}
\phi^\pm_{\text{on-sh.}}(x) =
\int (d^4 \vec{\bf k})  \, e^{\pm i kx} \, \delta(k^2 - m^2) a^\pm_{\text{on-sh.}} (k).
\end{eqnarray}
With the help of this representation we can readily make an extension of the considered fields beyond the on-shell surface.
To this goal, we introduce the following replacement
\begin{eqnarray}
\label{B-7}
\delta(k^2 - m^2) a^\pm_{\text{on-sh.}} (k) \Rightarrow
a^\pm_{\text{off-sh.}} (k),
\end{eqnarray}
where $a^\pm_{\text{off-sh.}} (k)$ denotes the off-shell field operator.
As a consequence, the off-shell commutation relation takes the form of
\begin{eqnarray}
\label{B-8}
 [ a^+_{\text{off-sh.}}(k), \, a^-_{\text{off-sh.}}(p)] = \frac{1}{4E^2(\vec{\bf k})}
 \delta^{(4)}(k - p)
\end{eqnarray}
which results in
\begin{eqnarray}
\label{B-9}
&&[\phi^+_{\text{off-sh.}}(x), \, \phi^-_{\text{off-sh.}}(y)] =
\nonumber\\
&&
\int \frac{(d^4 k)}{4E^2(\vec{\bf k})}
e^{ik(x - y)} \equiv D^+_{\text{off-sh.}}(x-y),
\end{eqnarray}
where $k_0 \not= E(\vec{\bf k})$ anymore.

%%%%%%%%%%%%%%%%%%%%%%%%
\section{On the Lorentz parametrization}
\label{NF-Par}
%%%%%%%%%%%%%%%%%%%%%%%%

In this Appendix, we illustrate schematically the new subtleties of Lorentz parametrization observed and 
discussed, in details, in \cite{Anikin:2021zxl, Anikin:2022eyf}.
For the pedagogical reason, we focus on the simplest hand-bag type of diagrams which describe the 
forward Compton scattering amplitude (CSA). The forward Compton amplitude reads
\begin{eqnarray}
\label{Amp-1}
{\cal A}_{\mu\nu} = \langle P| a^-_\nu(q) \, \mathbb{S}[\bar\psi, \psi, A] \, a^+_\mu(q) | P\rangle,
\end{eqnarray}
where the interaction $\mathbb{S}$-matrix is determined as
\begin{eqnarray}
\mathbb{S}[\psi,\bar\psi, A]=T\text{exp}\Big\{
i \int (d^4 z) \big[ {\cal L}_{QCD}(z) + {\cal L}_{QED}(z)\big] \Big\}.
\nonumber
\end{eqnarray}
We stress that in contrast to the photon Fock states the hadron states, see Eqn.~(\ref{Amp-1}),
cannot be expressed through the relevant operators of creation and annihilation.
Albeit, the creation and annihilation hadron operators can be introduced with the help of the effective Lagrangian
describing the transition of partons into hadrons, which is not our case.

Making used the commutation relations of creation (or annihilation) relevant operators with $\mathbb{S}$-matrix
together with Wick's theorem, we can readily derive the hand-bag diagram contribution to the CSA.
It reads (within the momentum representation)
 \begin{eqnarray}
\label{Amp-3}
&&{\cal A}_{\mu\nu}^{\text{hand-bag}} =
\int (d^4 k) \,\text{tr} \big[ E_{\mu\nu}(k)\Gamma  \big] \Phi^{[\Gamma]}(k),
\end{eqnarray}
where $\Gamma$ implies the corresponding $\gamma$-matrix and (here, $c$ stands for the connected diagram contributions)
\begin{eqnarray}
\label{E}
&&E_{\mu\nu}(k) = \gamma_\mu S(k+q) \gamma_\nu + \gamma_\nu S(k-q) \gamma_\mu,
\\
&&
\label{phifun}
\Phi^{[\Gamma]}(k)= \int (d^4 z) \, e^{ikz}
\langle P, S| T \bar\psi(0)\,\Gamma\, \psi(z) \mathbb{S}[\bar\psi, \psi, A] |P, S \rangle_c.
\end{eqnarray}
For the further discussion, it is instructive to introduce the notations as 
\begin{eqnarray}
\label{h-m-e-1}
&&\langle P, S| T \bar\psi(0)\,\Gamma\, \psi(z) \mathbb{S}[\bar\psi, \psi, A] |P, S \rangle \equiv 
\langle P, S| {\cal O}^{[\Gamma]} \big(0, z \,|\,  \bar\psi, \psi, A  \big) |P, S \rangle,
\nonumber\\
&&
\Phi^{[\Gamma]}(k)\stackrel{{\cal F}}{=} \langle P, S| {\cal O}^{[\Gamma]} \big(0, z \,|\,  \bar\psi, \psi, A  \big) |P, S \rangle,
\end{eqnarray}
where $\stackrel{{\cal F}}{=}$ denotes the corresponding Fourier transformations with the measure as  
\begin{eqnarray}
\label{ft}
d\mu(z)= (d^4 z) \, e^{ikz} \quad \text{or} \quad  d\mu(k)= (d^4 k) \, e^{- ikz}.
\end{eqnarray}

As discussed in \cite{Anikin:2022eyf},
$\mathbb{S}$-matrix generates the Wilson lines, which ensure the gauge invariance of non-local operators,
and the contributions which are not being exponentiated. The latter contributions provide {\it (a)} the
evolutions of the corresponding functions and {\it (b)} the tensor structure of Lorentz parametrization
(see Fig.~2 in \cite{Anikin:2022eyf}).
In the present paper, we mainly focus on the non-exponentiated contributions of $\mathbb{S}$-matrix
which do not refer to the Wilson lines. 

Notice that
the forward CSA we now consider corresponds to the simplest example from the viewpoint of factorization 
in comparison with the standard Drell-Yan-like processes.

Let us concentrate on the function $\Phi^{[\Gamma]}(k)$ which has to be parametrized based on the Lorentz covariance.
The hadron-to-hadron matrix element of different quark-gluon operators depends 
on the hadron (external) and quark-gluon (internal) vectors.  Namely, the hadron Fock states give 
the hadron momenta and spin (pseudo)vector dependences, while the quark-gluon operators, after the corresponding Fourier 
transformations, are forming the different combinations of parton momenta and spins.
That is, we have the following expressions,  here we use the parametrization of \cite{Bastami:2018xqd},
\begin{eqnarray}
\label{Me-D-1}
&&\langle P, S| {\cal O}^{[\gamma^+]} \big(0, z \,|\,  \bar\psi, \psi, A  \big) |P, S \rangle \Big |_{\mathbb{S}=\mathbb{I}}
=
\langle P, S|   \bar\psi(0)\,\gamma^+\, \psi(z) |P, S \rangle
\stackrel{{\cal F}}{=}
f_1 + \frac{\epsilon^{+- i j} k_\perp^i S^j_T}{m_N} f^\perp_{1\, T},
\\
&&
\langle P, S| {\cal O}^{[-i\sigma^{+i} \gamma^5]} \big(0, z \,|\,  \bar\psi, \psi, A  \big) |P, S \rangle \Big |_{\mathbb{S}=\mathbb{I}}=
\\
&&
\langle P, S|   \bar\psi(0)\,[-i\sigma^{+i} \gamma^5]\, \psi(z) |P, S \rangle
\stackrel{{\cal F}}{=}
S^i_T h_1 + S^i_L \frac{k^i_\perp}{m_N} h^\perp_{1\, L} + 
\frac{\kappa^{i j} S^j_T}{m^2_N} h^\perp_{1\, T} + 
\frac{\epsilon^{i j} k_\perp^j}{m_N} h_{1\, \perp}, \,\, \text{etc.}
\nonumber
\end{eqnarray}
where the condition of interaction absence, $\mathbb{S}=\mathbb{I}$, provides the standard Lorentz parametrizations.
Within this standard way, if we are interested in the corresponding evolutions of parton distributions,
the interaction, $\mathbb{S}\not=\mathbb{I}$, should be ``switched on'' after the Lorentz parametrizations. 

However, as explained in \cite{Anikin:2021zxl, Anikin:2022eyf}, the presence of interaction in the corresponding correlator
from the very beginning opens the window for the new parametrization functions. 
Indeed, if $\mathbb{S}\not=\mathbb{I}$, the quark-gluon operator, for example 
${\cal O}^{[\gamma^+]} \big(0, z \,|\,  \bar\psi, \psi, A  \big)$, has more complicated structure depending on the 
decomposition order over the coupling constant (here we use the notations of \cite{Anikin:2021zxl}):
\begin{eqnarray}
\label{Me-D-2}
{\cal O}^{[\gamma^+]} \big(0, z \,|\,  \bar\psi, \psi, A  \big) \Big|_{ \mathbb{S}_{QCD}^{(2)}} 
\stackrel{Fi. \, tr.}{
\Longrightarrow} 
\big[ \bar u^{(\up_x)}(k) \gamma^+ \gamma^\perp \gamma_5 u^{(\up_x)}(k)\big]
\big[ \bar u^{(\up_x)}(k) u^{(\up_x)}(k)\big] \sim s_\perp,
\end{eqnarray}
where $Fi. \, tr.$ denotes the implementation of the corresponding Fierz transformations (see \cite{Anikin:2021zxl} for all details).

The {\it r.h.s.} of Eqn.~(\ref{Me-D-2}) shows that, thanks to the interaction, 
one of the internal (quark) vectors can be associated with the quark spin
even if the hadron in the correlator is unpolarized. As a result, due to the interaction in the correlator, 
the parametrization of the vector projection, 
$\langle P, S| {\cal O}^{[\gamma^+]} \big(0, z \,|\,  \bar\psi, \psi, A  \big) |P, S \rangle$,
receives the addition parametrization functions:
\begin{eqnarray}
\label{Me-D-3}
&&\langle P, S| {\cal O}^{[\gamma^+]} \big(0, z \,|\,  \bar\psi, \psi, A  \big) |P, S \rangle \Big |_{\mathbb{S}\not=\mathbb{I}}
=
\\
&&
\langle P, S|  T \bar\psi(0)\,\gamma^+\, \psi(z)\,\,
\mathbb{S}^{(2)}[\bar\psi, \psi, A] |P, S \rangle\Big |_{\text{\tiny diag. with impl. loop integ.}}
\stackrel{{\cal F}}{=}
i \epsilon^{+ - P_\perp s_\perp} \tilde f_1^{(1)} + ....
\nonumber
\end{eqnarray}
Here, we do not show the other possible new functions which can be found in \cite{Anikin:2021zxl}.

To conclude this Appendix, we would like to notice that, in physical terms, 
the $k_\perp$-dependent function $\tilde f_1^{(1)}$ has been entirely generated by the 
quark spin alignment. From the mechanical point of view, it resembles the deviation of alike-rotated balls from 
the straightforward motion.

%%%%%%%%%%%%%%%%%%%%%%%%%%%%
\section{The covariant (invariant) integrations}
\label{AppC}
%%%%%%%%%%%%%%%%%%%%%%%%%%%%

In this appendix, we describe the method of the covariant (invariant) integration which is based on the
Lorentz decomposition. We begin with the simplest example which clearly illustrates the method.
We consider an arbitrary correlator with one Lorentz vector index which corresponds to the forward
hadron matrix element of the quark non-local operator. It reads
\begin{eqnarray}
\label{App-Inv-Int-1}
\langle P | \mathbb{O}_\alpha\big(\psi(0), \bar\psi(z)\big) | P \rangle
\stackrel{{\cal F}}{=}
\int (d^2 \vec{\bf k}_\perp) \, F(x, k_\perp ; P) \, k^\perp_\alpha \equiv J^\perp_\alpha,
\end{eqnarray}
where $\stackrel{{\cal F}}{=}$ denotes the Fourier transform and defined
by the integration measure as
\begin{eqnarray}
\label{F-t-1}
d\mu^-(k) = (dz^-) e^{-ik^+ \, z^-} \Big|_{k^+=x P^+}
\end{eqnarray}
and $a^\perp= (0, \vec{\bf a}_\perp, 0)$ for any four-vector $a$.
Here and in what follows, we omit the covariant and contravariant indices unless it leads to misunderstandings.
In Eqn.~(\ref{App-Inv-Int-1}), the hadron momentum $P$ represents the exterior Lorentz vector
characterizing the correlator $\langle P | \mathbb{O}_\alpha | P \rangle$
and we assume that $P=(P^+, 0^-, \vec{\bf P}_\perp)$.

We now perform the Lorentz decomposition of $J^\perp_\alpha$ over the exterior Lorentz vector,
we obtain that
\begin{eqnarray}
\label{App-Inv-Int-2}
J^\perp_\alpha = P^\perp_\alpha \, {\cal A} \equiv
P^\perp_\alpha \, \int (d^2 \vec{\bf k}_\perp) \, F(x, k_\perp ; P) \, \frac{(k^\perp \cdot P^\perp )}{P_\perp^2}.
\end{eqnarray}
The covariant spin axial-vector $S_\alpha$, which can form the Lorentz vector $i\varepsilon^{\alpha S_\perp + -}$,
is not participating in the decomposition of Eqn.~(\ref{App-Inv-Int-2}) even if the spin axial-vector characterizes the given
correlator. Indeed,  if $S=(\lambda P^+/M, 0^-, \vec{\bf S}_\perp)$ then $(P\cdot S) = (P_\perp\cdot S_\perp)=0$ and,
hence,  $i\varepsilon^{P_\perp S_\perp + -}\not= 0$. So, the Lorentz vector $i\varepsilon^{\alpha S_\perp + -}$
is not an orthogonal basis vector for the decomposition.

The Eqns.~(\ref{App-Inv-Int-1}) and (\ref{App-Inv-Int-2})
correspond, roughly speaking, to two different vector representations of the single Lorentz vector $J^\perp_\alpha$.
Let us focus on the scalar product $(J^\perp\cdot B^\perp)$ where an arbitrary exterior vector $B$ which is not associated with the
correlator of Eqn.~(\ref{App-Inv-Int-1}). We have the following expression
\begin{eqnarray}
\label{App-Inv-Int-3}
&&(J^\perp\cdot B^\perp) =
 \int (d^2 \vec{\bf k}_\perp) \, F(x, k_\perp ; P) \, (k^\perp \cdot B^\perp) =
 \nonumber\\
 &&
(P^\perp\cdot B^\perp) \, \int (d^2 \vec{\bf k}_\perp) \, F(x, k_\perp ; P) \, \frac{(k^\perp \cdot P^\perp)}{P_\perp^2}
\end{eqnarray}
which involves $J^\perp_\alpha$ written before (see the first line of Eqn.~(\ref{App-Inv-Int-3})) and after (see the
second line of Eqn.~(\ref{App-Inv-Int-3})) implementation of the covariant (invariant) integration.

%
%
%%%%%%%%%%%%%%%%%%%%%%%%%%%%%%% FIGURE %%%%%%%%%%%%%%%%%%%%%%%%%%%%%%%%
\begin{figure}[tbp]
\centering % \begin{center}/\end{center} takes some additional vertical space
\includegraphics[width=.6\textwidth]{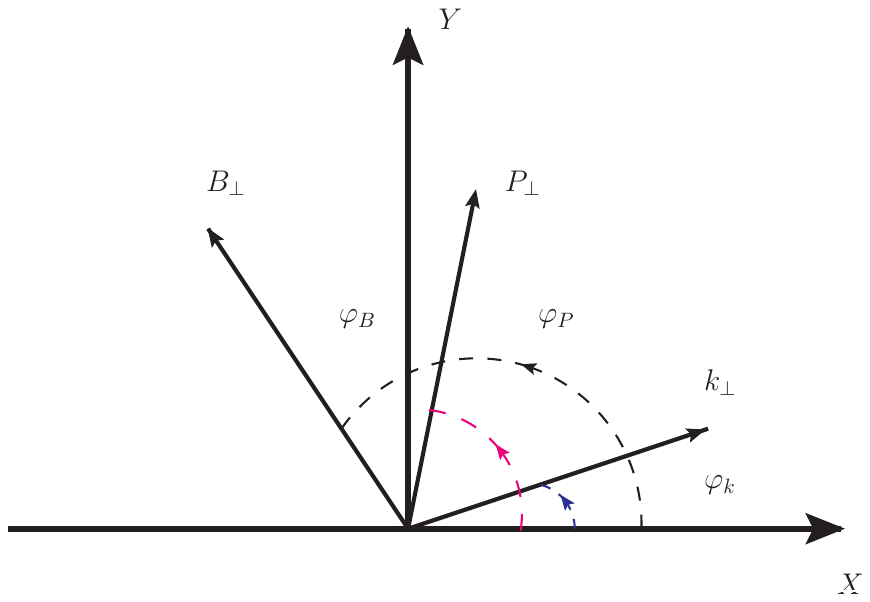}
% "\includegraphics" is very powerful; the graphicx package is already loaded
\vspace{-6cm}
\caption{\label{Fig-Ap-1} The definitions of angles $\varphi_B$, $\varphi_P$ and $\varphi_k$.}
\end{figure}
%%%%%%%%%%%%%%%%%%%%%%%%%%%%%%%%%%%%%%%%%%%%%%%%%%%%%%%%%%%%%%%%%%%%%%%
%
%%%%%%%%%

From  Eqn.~(\ref{App-Inv-Int-3}), we can immediately derive the necessary condition for the self-consistency of the Lorentz decomposition
of Eqn.~(\ref{App-Inv-Int-2}).
Indeed, Eqn.~(\ref{App-Inv-Int-3}) can be equivalently rewritten as
\begin{eqnarray}
\label{App-Inv-Int-4}
&& \int (d^2 \vec{\bf k}_\perp) \, F(x, k_\perp ; P) \,
 \Big \{ (k^\perp \cdot B^\perp) -
(P^\perp\cdot B^\perp) \,  \frac{(k^\perp \cdot P^\perp)}{P_\perp^2} \Big\} =
\nonumber\\
&&
\int (d^2 \vec{\bf k}_\perp) \, F(x, k_\perp ; P)  | k_\perp| \,|B^\perp| \,
\Big \{ \cos\varphi_{Bk} - \cos\varphi_{BP} \cos\varphi_{Pk} \Big\}=0,
\end{eqnarray}
where the notation $\varphi_{\mathbb{A} \mathbb{B}} =
\varphi_{\mathbb{A}} - \varphi_{\mathbb{B}}$ with $(\mathbb{A},\mathbb{B})\equiv (B_\perp, P_\perp, k_\perp)$
has been used for the corresponding angles, see Fig.~\ref{Fig-Ap-1}.

One can readily solve Eqn.~(\ref{App-Inv-Int-4}), we have the solution given by ($\varphi_{Bk} = \varphi_{BP} + \varphi_{Bk}$)
\begin{eqnarray}
\label{Sol-1-App}
&&\varphi_{B}=\varphi_{P} \pm n\pi \,,
\\
&&
\label{Sol-1-App-2}
\varphi_{P}=\varphi_{k} \pm n\pi.
\end{eqnarray}
These conditions ensure the self-consistency of the covariant integration (see  Eqn.~(\ref{App-Inv-Int-1})).
Moreover, the (anti)collinearity of $P_\perp$ and $B_\perp$, see Eqn.~(\ref{Sol-1-App}),
is a necessary and enough condition for the mentioned self-consistency, while
the condition given by Eqn.~(\ref{Sol-1-App-2}) is not crucial and can be omitted.

We now dwell on the practical example of the covariant (invariant) integration applied to Eqn.~(\ref{V-me}).
In this case, we have the following decomposition
\begin{eqnarray}
\label{App-Inv-Int-5}
K^\perp_\alpha\equiv\int (d^2 \vec{\bf k}_\perp) \, f_{(2)}(x, k_\perp, s_\perp ; P) \, k^\perp_\alpha =
P^\perp_\alpha \, {\cal B}  +
\varepsilon^{\alpha s_\perp + -} {\cal C}
\end{eqnarray}
where $s_\perp$ implies the covariant quark spin vector and
\begin{eqnarray}
\label{B-App}
&&
{\cal B}=\int (d^2 \vec{\bf k}_\perp) \, f_{(2)}(x, k_\perp, s_\perp ; P) \, \frac{(k^\perp \cdot P^\perp)}{P^2_\perp},
\\
&&
\label{C-App}
{\cal C}=\int (d^2 \vec{\bf k}_\perp) \, f_{(2)}(x, k_\perp, s_\perp ; P) \, \frac{\varepsilon^{k_\perp s_\perp + -}}{s^2_\perp}.
\end{eqnarray}
Notice that $\varepsilon^{k_\perp s_\perp + -}\not= 0$ otherwise we deal with a trivial case of ${\cal C}=0$
provided $s^2_\perp\not= 0$.
The important requirement for the Lorentz decomposition of Eqn.~(\ref{App-Inv-Int-5}) is
that the vectors $P^\perp$ and $\varepsilon^{\alpha s_\perp + -}$ have to be orthogonal ones, {\it i.e.} they are
forming the orthogonal system. In this context, we have to derive the necessary conditions for the orthogonality.

Consider the scalar product $(P_\perp \cdot K^\perp )$ giving
\begin{eqnarray}
\label{App-Inv-Int-6}
(P_\perp \cdot K^\perp ) =
P^2_\perp \, {\cal B}  +
\varepsilon^{P_\perp s_\perp + -} {\cal C}
\end{eqnarray}
which takes place if and only if
\begin{eqnarray}
\label{App-Req-1}
\varepsilon^{P_\perp s_\perp + -}\sim \sin\varphi_{Ps} =0 \Rightarrow \varphi_P = \varphi_s \pm n\pi.
\end{eqnarray}
That is, the vectors $P^\perp$ and $s_\perp$ have to be collinear (or anti-collinear) ones.
On the other hand, one may contract $K^\perp_\alpha$ in (\ref{App-Inv-Int-5}), with the vector $\varepsilon^{\alpha s_\perp + -}$.
This case also leads to the condition of Eqn.~(\ref{App-Req-1}) .

To conclude, the condition of Eqn.~(\ref{App-Req-1}) is the necessary condition for the Lorentz decomposition given by
Eqn.~(\ref{App-Inv-Int-5}).

%%%%%%%%%%%%%%%%%%%%%%%%%%%%%%%%%%%%%%%%%%%%%%%%%%%%%%%%%%%%%%%%%%%%%%%%%%%%%
%%%%%%%%%%%%%%%%%%%%%%%%%%%%%   References   %%%%%%%%%%%%%%%%%%%%%%%%%%%%%%%%%%
%%%%%%%%%%%%%%%%%%%%%%%%%%%%%%%%%%%%%%%%%%%%%%%%%%%%%%%%%%%%%%%%%%%%%%%%%%%%%

\end{document}